\documentclass[11pt]{article}

\usepackage{chemformula} % Formula subscripts using \ch{}
\usepackage[T1]{fontenc} % Use modern font encodings

\usepackage{amsmath, amssymb,mathtools}
\usepackage{siunitx}

\usepackage{graphicx}
\usepackage{subcaption}
\usepackage{booktabs}
\usepackage{tikz-cd}

\usepackage{xcolor}
\usepackage[prefix=sol-]{xcolor-solarized}

\usepackage{hyperref}
\usepackage{authblk}
\usepackage{comment}
\usepackage{mod}
\usepackage{listings}

\author[1,2]{Erika M. Herrera Machado}
\author[1]{Jakob L. Andersen}
\author[1]{Rolf Fagerberg}
\author[3,1]{Daniel Merkle}

\newcommand\email[1]{\texttt{#1}}

\affil[1]{Department of Mathematics and Computer Science,
    University of Southern Denmark, Odense, Denmark
    \email{\{jlandersen,rolf,machado\}@imada.sdu.dk}}
\affil[2]{Faculty of Mathematics and Computer Science, Friedrich Schiller University Jena, Jena, Germany}
\affil[3]{Algorithmic Cheminformatics Group, Faculty of Technology \& Center for Biotechnology (CeBiTec),
    Bielefeld University, Bielefeld, Germany
    \email{daniel.merkle@uni-bielefeld.de}}

\title{A Sensitivity Analysis Methodology for Rule-Based Stochastic Chemical Systems}

\begin{document}
\maketitle	

%%%%%%%%%%%%%%%%%%%%%%%%%%%%%%%%%%%%%%%%%%%%%%%%%%%%%%%%%%%%%%%%%%%%%

\begin{abstract}
    In this study, we introduce a sensitivity analysis methodology for
    stochastic systems in chemistry, where dynamics are often governed by
    random
    processes. Our approach is based on gradient estimation via finite
    differences,
    averaging simulation outcomes, and analyzing variability under intrinsic
    noise.
    We characterize gradient uncertainty as an angular range within which all
    plausible gradient directions are expected to lie. A key feature of our approach is that this uncertainty measure
    adaptively guides the number of simulations performed for each
    nominal-perturbation pair of points in order to minimize unnecessary
    computations while maintaining robustness.

    Systematically exploring a range of parameter values across the parameter
    space, rather than focusing on a single value, allows us to identify not
    only
    sensitive parameters but also regions of parameter space associated with
    different levels of sensitivity. These results are visualized through
    vector
    field plots to offer an intuitive representation of local sensitivity
    across
    parameter space. Additionally, global sensitivity coefficients over sampled points
    in the parameter space are computed
    to
    capture overall trends.

    Flexibility regarding the choice of output observable measures is another
    key feature of our method: while traditional sensitivity analyses often
    focus
    on species concentrations, our framework allows for the definition of a
    large
    range of problem-specific observables. This makes it broadly applicable in
    diverse chemical and biochemical scenarios. We demonstrate our approach on
    two
    systems: classical Michaelis-Menten kinetics and a rule-based model of the
    formose reaction, using the cheminformatics software MØD for
    Gillespie-based
    stochastic simulations. 
\end{abstract}

%%%%%%%%%%%%%%%%%%%%%%%%%%%%%%%%%%%%%%%%%%%%%%%%%%%%%%%%%%%%%%%%%%%%%%%%%%%%%%%
\section{Introduction}\label{Sec:Intro}
%%%%%%%%%%%%%%%%%%%%%%%%%%%%%%%%%%%%%%%%%%%%%%%%%%%%%%%%%%%%%%%%%%%%%%%%%%%%%%%

Mathematical modeling has become increasingly important for understanding
biological and chemical systems. These models not only serve as predictive
tools but also offer valuable insights into the
mechanisms underlying observed behaviors. Common model parameters include
reaction rates, initial conditions, activation energies, and thermodynamic
constants. Estimating the values of such parameters is often subject to
uncertainty~\cite{saltelli2005sensitivity, rabitz1983sensitivity}.

Sensitivity analysis is the study of how input parameters influence the model
output. If small changes in a parameter result in significant variations in the
outcome, the model is considered sensitive to that parameter. The application
of sensitivity analysis in chemical and biochemical systems serves multiple
purposes, including: i) identifying the key parameters in the model, ii)
understanding how changes in input values influence the model output, iii)
assessing the robustness of the model, iv) simplifying the model by removing
parameters of low impact, and v) inferring the model's output in the vicinity
of a given point without the need to rerun the
experiment~\cite{saltelli2005sensitivity, rabitz1983sensitivity,
    plyasunov2007efficient, gunawan2005sensitivity, turanyi1990sensitivity}. We
elaborate on the background of the field in section~\nameref{Sec:RW}.

For modeling the time evolution of chemical systems, a stochastic approach is
often more suitable than a deterministic one, especially for complex systems,
systems prone to fluctuations, or systems where discreteness and stochasticity
play important roles~\cite{rabitz1983sensitivity, turanyi1990sensitivity,
    bartholomay1958stochastic, gillespie1976general}. However, the uncertainty
and
noise present in stochastic systems add additional challenges to sensitivity
analysis. Section~\nameref{Sec:SCS} briefly introduces stochastic chemical
systems and modeling formalisms.

In this work, we propose a sensitivity analysis methodology for discrete
stochastic systems. A major strength of our proposal is its flexibility in
defining the subject of the analysis, which may be any observable that the user
can define on the simulation trace\footnote{Simulation trace: a record of the
    time evolution of a system during a single simulation run.}. As examples,
an
observable can be the number of molecular species~$A$ present at the end, the
time elapsed before molecular species~$B$ appears for the first time, the
number of times reaction~$R$ took place before reaction~$S$ first occurred, or
the average number of carbons per molecule in the molecular population at the
end. Once an objective has been chosen, we then average its value over repeated
runs of the stochastic simulation to account for intrinsic noise.
The next ingredient is to estimate the gradient of the output observable
function using finite differences. We do that not just at a single point in parameter space (which in this paper we refer to as \textit{nominal point}), but across a set of values for each parameter, that is, for a set of nominal points. This exploration of a
range of values allows us to detect not only sensitive parameters but also
specific regions associated with high or low sensitivity.

A key feature of our method is that the number of repetitions of each
simulation is chosen in a structured manner that aims at balancing
computational time with precision achieved at each point in the parameter
space. For this, we leverage statistical concepts such as the Central Limit
Theorem for quantifying the uncertainty of the estimated gradient as an angular
uncertainty of the gradient's direction.
Since this uncertainty varies for each nominal-perturbation pair of points, we
adapt the number of simulations accordingly to ensure consistent precision.
Finite difference approaches in the literature have proven to be
effective~\cite{rabitz1983sensitivity, turanyi1990sensitivity,
    gunawan2005sensitivity, rathinam2010efficient}. However, some works argue
against it due to the excessive number of function evaluations required and its
unreliability in the presence of
noise~\cite{nocedal1999numerical,saltelli2005sensitivity}. Working with
low-dimensional parameter spaces in rule-based systems (which will be introduced in subsection \nameref{Subsec:Rule-based}) and employing a
systematic and adaptive method for result averaging help us mitigate these
issues in our approach.

Additionally, we present results as a series of vector fields over the
parameter space, offering a more interpretable view compared to other
sensitivity analysis techniques. These visualizations help identify directions
within the input space for improving robustness or optimizing observable
values, and capture the combined effects of multiple parameters. We also
calculate global sensitivity coefficients for each parameter to quantify their impact
on the model and importance across sampled points in the parameter space.

The rest of this paper is structured as follows. In section \nameref{Sec:SCS},
we introduce stochastic chemical systems and modeling formalisms.
\nameref{Sec:RW} provides a review of existing approaches for deterministic and
stochastic modeling. In section \nameref{Sec:Method}, we describe our approach
in detail. \nameref{Sec:Examples} illustrates this methodology using the
well-known Michaelis-Menten kinetics and a rule-based model for formose
chemistry. Finally, in \nameref{Sec:Conclusion} we close the paper by
summarizing the work we have carried out.

%%%%%%%%%%%%%%%%%%%%%%%%%%%%%%%%%%%%%%%%%%%%%%%%%%%%%%%%%%%%%%%%%%%%%%%%%%%%%%%
\section{Simulation of Stochastic Chemical Systems}\label{Sec:SCS}
%%%%%%%%%%%%%%%%%%%%%%%%%%%%%%%%%%%%%%%%%%%%%%%%%%%%%%%%%%%%%%%%%%%%%%%%%%%%%%%

Stochastic chemical systems describe the behavior of chemical reactions where
the evolution of molecular populations is inherently random. Such systems are
particularly prominent in biological and biochemical contexts, where molecular
populations can be small, and random fluctuations play a significant role in
system behavior. For example, gene expression, cellular signaling, and
metabolic pathways often exhibit stochastic dynamics that cannot be captured by
deterministic models~\cite{gunawan2005sensitivity, rathinam2010efficient,
    komorowski2011sensitivity}.

Unlike deterministic models, which predict exact outcomes based on initial
conditions, stochastic models account for this probabilistic nature of
molecular interactions~\cite{rabitz1983sensitivity, turanyi1990sensitivity,
    bartholomay1958stochastic, gillespie1976general}. Stochastic chemical
systems
are typically modeled as continuous-time Markov
processes~\cite{gunawan2005sensitivity, rathinam2010efficient,
    komorowski2011sensitivity} where the state of the system is represented by
a
vector of non-negative integers corresponding to the number of molecules of
each species. The system evolves through discrete reaction events, each
characterized by a propensity function that defines the probability of a
reaction occurring in a given time interval. The evolution of the system’s
probability distribution over time is described by the Chemical Master Equation
(CME), a differential equation that describes the time-dependent probability of
the system being at each possible state. The CME models stochastic dynamics of
reacting systems accounting for the inherent randomness in molecular events.
However, the CME is often intractable to solve analytically. Instead, numerical
trajectories of the system can be efficiently generated using stochastic
simulation algorithms (SSA), such as the Direct Method introduced by
Gillespie~\cite{gillespie1976general}, in which for each reaction $r$ a
propensity function is defined which measures the probability that $r$ will
occur in the system in the next time interval $[t, t+dt)$. 

\subsection{Rule-Based Modeling}\label{Subsec:Rule-based}

For systems with
complex reaction mechanisms involving numerous molecular interactions, generative
rule-based modeling provides a powerful alternative to explicitly enumerating
all possible reactions~\cite{danos2007rule, chylek2014rule, benko2003graph}. This paradigm addresses the
potential combinatorial complexity associated with these chemical systems by
representing molecules as agents and interactions as rules that describe how
local patterns should be modified. %This way, it is possible to avoid the need for enumerating all possible reactions between all possible molecular species~\cite{danos2007rule, benko2003graph}. 
In other words, rule-based
modeling offers a powerful simplification of complex biochemical systems by
describing them with a concise set of rules rather than an extensive set of
specific reactions. Furthermore, being generative, the system can be
initialized with a small number of molecular species instead of a comprehensive
list of possible species. This abstraction is especially advantageous for
sensitivity analysis: by naturally reducing the dimensionality of the parameter
space, rule-based models make the analysis more tractable, both computationally
and conceptually. In the context of our approach, this allows for a more
efficient exploration of system behavior and facilitates the visualization of
the sensitivity across the parameter space.
One computational framework working in this paradigm is the software package
MØD~\cite{andersen2016software}, which approaches rule-based generative
chemistry via graph grammars where atoms are explicitly represented. MØD
contains a module for performing Gillespie-based stochastic
simulation~\cite{machado2025rulebasedgillespiesimulationchemical}, which forms
the vehicle by which we display our methodology. This module uses reaction rules to dynamically generate the relevant parts of the underlying
reaction networks during the simulation, capturing the stochastic evolution of
molecular populations while significantly reducing computational complexity.
%This is particularly useful for modeling biochemical networks with large numbers of interacting species, where traditional reaction-based representations become infeasible.

%%%%%%%%%%%%%%%%%%%%%%%%%%%%%%%%%%%%%%%%%%%%%%%%%%%%%%%%%%%%%%%%%%%%%%%%%%%%%%%
\section{Related Work}\label{Sec:RW}
%%%%%%%%%%%%%%%%%%%%%%%%%%%%%%%%%%%%%%%%%%%%%%%%%%%%%%%%%%%%%%%%%%%%%%%%%%%%%%%

\subsection{Previous Work on Sensitivity Analysis in Chemical Systems}

In the context of chemical kinetics, categorizing methodologies is a
multidimensional task. In this paper, we consider the following dimensions for
discussing and categorizing existing work:
\begin{itemize}
    \item Scope: global (evaluation of parameter importance across the entire input space) or local (evaluation of model response to small perturbations around a single nominal point).
    \item Modeling formalism: traditional ODE system,
          %CME for stochastic systems, 
          or Gillespie SSA-based.
    \item Source of randomness: the input variable values, or intrinsic
          stochastic effects.
    \item Output observable of the analysis: species concentrations are the
          standard, but other interesting measures can be considered.
    \item Time handling: as a variable, or fixed in advance.
\end{itemize}

In the following, we highlight mainly the works that provide the necessary
context for our study, in order to illustrate how it fits into this broad and
evolving field. The selected citations are organized to reflect the historical
trend of chemical kinetics formalisms from deterministic ODE-based models to
the adoption of Gillespie SSA-based techniques for stochastic systems.

\subsubsection{Deterministic Modeling}

Traditional sensitivity analysis is applied to continuous deterministic
systems. The evolution of a spatially homogeneous chemical system is typically
modeled by the ordinary differential equation system
\begin{equation}\label{eq:det_evol}
    \frac{d\mathbf{c}}{dt} = \mathbf{f}(\mathbf{c,x},t), \quad
    \mathbf{c}(0)=\mathbf{c}^{0},
\end{equation}
where $\mathbf{c}$ is the vector of species concentrations, $\mathbf{x}$ is the
vector of system parameters, $t$ represents time, and $\mathbf{c}^{0}$ is the
vector of initial conditions.

Local methods generally produce sensitivity coefficients representing
parametric gradients at single points in the parameter
space~\cite{rabitz1983sensitivity, turanyi1990sensitivity, morio2011global}. We
want to highlight two methods that are regarded as the simplest and likely the
most intuitive in this category. The first one, assuming Eq.~\ref{eq:det_evol}
is solved for various sets of parameter values, consists of calculating local
first-order sensitivity coefficients with respect to the $j$th parameter $x_j$,
defined as
\begin{equation}
    \frac{\partial \mathbf{c}(t)}{\partial x_j}.
\end{equation}
These coefficients inform us about the effect on $\mathbf{c}(t)$ of small variations in the
parameter around its nominal value. The rest of parameters
are fixed, therefore this approach falls in the class of the One-at-a-Time
(OAT) methods. Higher-order coefficients can be obtained from the Taylor series
expansion.
Finite differences are commonly used to approximate these
gradients~\cite{rabitz1983sensitivity, turanyi1990sensitivity}, this approach
is also known as brute-force sensitivity analysis. The next classical method is
referred to as the direct method, as it involves the solution of the
differential equations that govern the sensitivity coefficients. In other
words, directly differentiating Eq.~\ref{eq:det_evol} with respect to $x_j$ we
obtain
\begin{equation}
    \frac{d}{dt}\frac{\partial\mathbf{c}}{\partial x_j} =
    \mathbf{J}(t)\frac{\partial\mathbf{c}}{\partial x_j} +
    \frac{\partial\mathbf{f}(t)}{\partial x_j},
\end{equation}
where $\mathbf{J}(t)=\frac{\partial\mathbf{f}}{\partial \mathbf{c}}$ and the
initial condition for $\frac{\partial\mathbf{c}}{\partial x_j}$ is a zero
vector. We refer to Refs.~\cite{rabitz1983sensitivity,
    turanyi1990sensitivity} for a more comprehensive review of local methods
and
solutions to the equations discussed.

Global methods are useful when one considers large variations of the system
parameters. As pointed out in Ref.~\cite{rabitz1983sensitivity}, the most
intuitive approach would consist of generating a solution surface in the
parameter space by calculating solutions of Eq.~\ref{eq:det_evol} for different
parameter values. However, this task can become unfeasible for a large number
of species and parameters. Two of the foundational global methods that we will
briefly introduce here rely on the assumption that the parameter vector
$\mathbf{x}$ is a random vector with known (or somehow estimable) probability
distribution. The first one of them is proposed by
Ref.~\cite{costanza1981stochastic}, where the task of sensitivity analysis
consists of determining the probability distribution $p$ of the concentrations
$\mathbf{c}(t)$, given the probability density function (pdf) of the
parameters. They achieve this by solving the partial differential equation
\begin{equation}
    \frac{\partial p}{\partial t} = \nabla (Fp),
\end{equation}
where $F = \frac{d\mathbf{z}}{dt}$, with $\mathbf{z}$ being the joint vector of
concentrations and parameters. In this context, the authors define stochastic
sensitivity coefficients as $\frac{\partial p}{\partial x_j}$. The next global
method is the Fourier amplitude sensitivity test
(FAST)~\cite{cukier1978nonlinear, cukier1975study}. Here, the model parameters
are expressed as periodic functions of a search variable $s$. This is done
using transformations of the form:
\begin{equation}
    x_j = G_j \left( \sin\left(\omega_j s\right) \right).
\end{equation}

This reformulation converts a multidimensional integral over the parameters
into a one-dimensional integral over $s$. The transformation functions $G_j$
are uniquely determined (see references for details). By selecting appropriate
integer frequencies $\omega_j$, the model outputs become $2\pi$-periodic
functions of $s$ and can be analyzed using Fourier methods.

Other traditional methods include: i) Monte Carlo methods, where parameter
values are generated based on the probability distributions, then solving the
system and analyzing results statistically~\cite{saltelli2005sensitivity}, ii)
the Morris screening method~\cite{morris1991factorial, saltelli2005sensitivity,
    zador2006local} for identifying influential parameters, and iii)
variance-based
methods such as Sobol indices~\cite{sobol1990sensitivity}, which quantify the
contribution of input parameters to output variability.

The methods presented above are historically among the foundational ones in the
field of sensitivity analysis for studying chemical kinetics. We have already
categorized them along the local/global dimension and now look at the other
categorical dimensions. For all these methods, their modeling formalism is the
traditional deterministic view of chemical kinetics where species
concentrations over time are modeled through a set of ODEs.
The term “stochastic” earlier did occur, but then
generally~\cite{atherton1975statistical, miller1983sensitivity} referred to
uncertainty in the parameter vector values, i.e., how variability in the
parameter vector $\mathbf{x}$ generates uncertainty in the concentration vector
$\mathbf{c}(t)$. However, some works already approached stochastic systems
where the randomness comes instead from the molecular events, especially in
systems with a small amount of molecules. Stochastic differential equations
were still a standard to handle these cases, usually by incorporating noise
terms.
Regarding the output of the analysis, the general tendency was the
concentrations of the molecular species. In the methods presented above, time
has been handled as an explicit independent variable to analyze the system's
evolution over time.

\subsubsection{Stochastic Modeling}

The next block of citations focuses on those where Gillespie's SSA-based
techniques are employed to model stochastic kinetics and where the source of
randomness comes from the internal stochastic effects due to the probabilistic
nature of biochemical reactions, rather than variability in parameter values
themselves.

In Ref.~\cite{gunawan2005sensitivity}, the authors present a methodology for
parametric sensitivity analysis based on density function sensitivity for
discrete stochastic systems, modeled by CMEs. Deriving sensitivity coefficients
directly from the CME is complex, so the authors estimate the pdf of the system
by generating SSA realizations of its states over time. They then derive
sensitivity measures employing finite differences and the Fisher Information
Matrix (FIM) to analyze how changes in the input parameters affect the pdf.
Their analysis is applied to the Schlögl model and a synthetic genetic
toggle-switch model. A discussion between stochastic and deterministic analyses
is provided as well.

Ref.~\cite{degasperi2008sensitivity} explores both local and global
sensitivity analysis methods applied to biochemical models, comparing
deterministic and stochastic approaches to SA. The analysis output is species
concentrations evaluated at specific fixed time points. Local OAT methods are
adapted for stochastic models using SSA, in one approach averaging simulation
outputs, and in another employing histogram distances to quantify differences
in the output distributions. For global analysis, a variance-based method is
discussed, which also utilizes histogram distances. They analyze the Schlögl
model and the MAPK signaling pathway.

In Ref.~\cite{rathinam2010efficient}, the authors investigate sensitivity
via finite perturbations in stochastic chemical systems using SSA simulations.
They introduce two methods for reducing the variance in the statistical
estimator described in Ref.~\cite{gunawan2005sensitivity}: the Common Random
Numbers (CRN) and Common Reaction Path (CRP). CRN reduces variance by
introducing dependence between perturbed and unperturbed simulations through
shared random number streams. CRP extends this approach by ensuring tighter
coupling between reaction paths via independent streams per reaction channel,
leading to more continuous perturbations and lower variance compared to CRN.
Comparatively, CRP outperforms CRN in terms of variance reduction and estimator
efficiency.

In Ref.~\cite{sheppard2013spsens}, SPSens is introduced as a software
package for efficient computation in stochastic parameter sensitivity analysis
of biochemical reaction networks. The software estimates sensitivity
coefficients related to the expected value of the system output to account for
the stochastic nature of the systems considered. These coefficients are
estimated using SSA realizations. Several sensitivity algorithms are
implemented in C, including the finite difference approach and the
aforementioned CRN and CRP, among others. It also includes variance reduction
methods for improved precision.

Ref.~\cite{damiani2013parameter} presents a numerical methodology for
parameter sensitivity analysis in catalytic reaction networks, focusing on how
parameters influence the probability distribution of the model output. Their
sensitivity measure is based on the one developed by
Ref.~\cite{gunawan2005sensitivity}, with the difference that here, the
sensitivity coefficients, rather than referring to all the output variables
described by a single probability distribution, analyze sensitivity with
respect to custom individual variables. They explore a range of parameter
values rather than focusing on a single nominal value, which results in a more
informative sensitivity assessment. SSA is used to simulate the stochastic
dynamics, and a kernel method approximates the pdfs of the output variables.
The parameters are ranked based on their influence.

The authors in Ref.~\cite{morshed2017efficient} review established
finite-difference estimators for sensitivity analysis, such as the previously
mentioned CRP and CRN from Ref.~\cite{rathinam2010efficient}, and the
Coupled Finite Difference (CFD) method by Ref.~\cite{anderson2012efficient}.
They propose a novel strategy for application to stiff systems, coupling paths
that are obtained via tau-leaping, and show significant improvement in
efficiency and similar accuracy to the CFD method.

In the domain of sensitivity analysis for stochastic chemical/biochemical
systems, the cited works show that the use of Gillespie's SSA to account for
intrinsic stochastic effects is by now standard practice. These approaches
often focus on estimating pdfs or sensitivity coefficients to understand the
system behavior under slight perturbations in the input parameter values, which
are generally fixed in advance. The scope of the reviewed papers is primarily
local, as they analyze effects around fixed nominal parameter values, although
some studies explore a broad range of parameter values as in
Ref.~\cite{damiani2013parameter}. Regarding the observables of the analysis,
species concentrations are predominant. Time handling is varied; some
approaches conduct the analyses at fixed, predetermined time points, while
other methods treat time as a variable, where sensitivity is assessed over
entire time trajectories.

The reviewed methods significantly contribute to understanding sensitivity in
stochastic chemical kinetics. However, some often occurring limitations
include: relying on a fixed nominal value per parameter, rather than a range of
values; focusing predominantly on species concentrations as observables; or
fixing in advance the number of simulations required. In our approach, we
overcome these limitations by proposing a methodology that incorporates
intrinsic stochasticity via Gillespie simulations and provides a geometric
characterization of gradient uncertainty. The parameter space is explored for a
range of values for each individual parameter, adapting the number of
simulations based on the precision of the gradient estimate. Additionally, the
choice of output observables is very flexible. We aim to offer an efficient,
robust, and adaptable framework for sensitivity analysis of chemical systems
where results are intuitive and visual.

%%%%%%%%%%%%%%%%%%%%%%%%%%%%%%%%%%%%%%%%%%%%%%%%%%%%%%%%%%%%%%%%%%%%%%%%%%%%%%%
\section{Methodology}\label{Sec:Method}
%%%%%%%%%%%%%%%%%%%%%%%%%%%%%%%%%%%%%%%%%%%%%%%%%%%%%%%%%%%%%%%%%%%%%%%%%%%%%%%

\subsection{Overview}\label{subsec:Overview}

We consider sensitivity analysis of a stochastic model with $n$ real-valued
input parameters of interest. This could for instance be a chemical reaction
network with $n$ of the reaction rate constants as parameters.
For sensitivity analysis to be meaningful, it is important to explore a
representative range of values for each parameter~\cite{damiani2013parameter},
since a system's observable may exhibit low sensitivity to changes in a
parameter around one reference value, but high sensitivity around another.
We therefore allow the user to define a subset $X \subseteq \mathbb{R}^n$ of
the $n$-dimensional Euclidean parameter space, which will be the parameter
values studied. The observable can be any user-defined value computable from
the simulation trace (i.e., the sequence of system states and events during one
simulation).

We make the assumption on the stochastic system and the observable chosen that
if the same simulation for a fixed parameter vector $\mathbf{x}\in X$ is
repeated many times, the average of the observed values converges to some
value. Thinking of this value as a function of $\mathbf{x}$, our goal will be
to estimate the gradient of that function. For this, we use the method of
finite differences, i.e. for a given nominal point $\mathbf{x}\in X$, perturb
one of its components slightly, and find the average observable value of
repeated simulations on both $\mathbf{x}$ and on the perturbed point. Then use
these averages to find an approximation to the corresponding component of the
gradient vector.

A key element of our methodology is to choose the number of repetitions in an
adaptive fashion:
Given a nominal point $\mathbf{x}\in X$ and a perturbed point $\mathbf{x'}$, we
run the stochastic simulation algorithm an initial number of times $N$,
independently for each. From the resulting collection of output traces, we
extract the corresponding observable values $\{y_1, \ldots, y_{N}\}$ for
$\mathbf{x}$ and $\{y_1', \ldots, y_{N}'\}$ for $\mathbf{x'}$.
We then continue to do further repetitions of the simulations until one of two
stopping criteria is met: either i) the runtime limit is reached, or ii) we
have achieved a desired \textit{gradient-uncertainty angle} (to be defined
below). In both cases, the averages of the collections $\{y_i\}_{i \in
    \mathbb{N}}$,  $\{y_i'\}_{i \in \mathbb{N}}$ are returned along with other
parameters of interest.

\begin{figure}
    \centering
    \includegraphics[width=9cm]{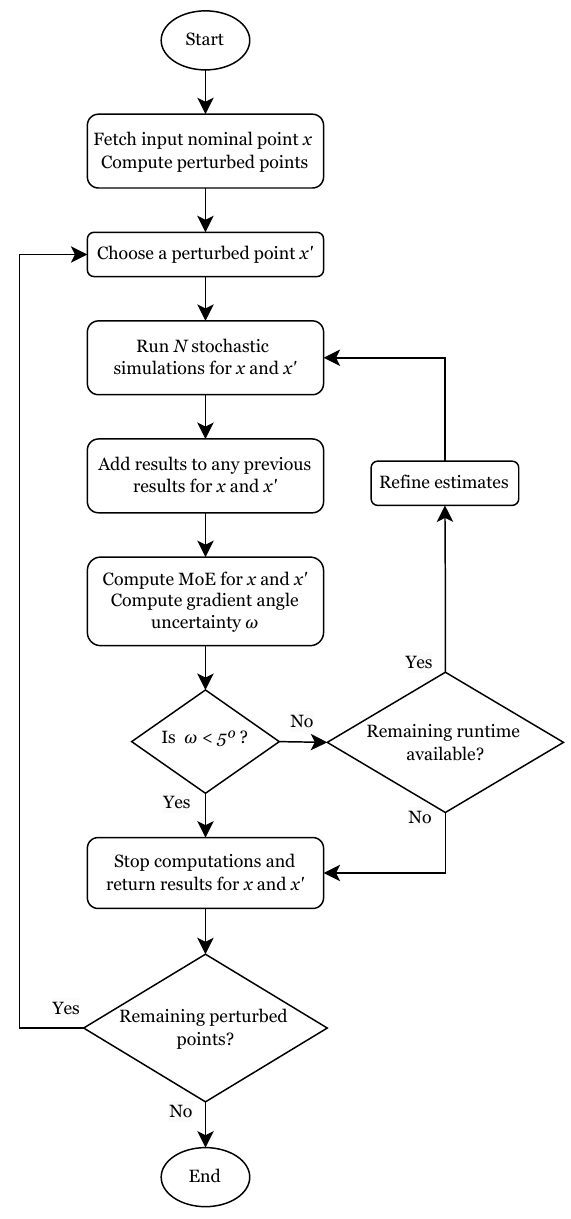}
    \caption{Data Collection Workflow. For each nominal–perturbation pair, we
        first
        run a batch of $N$ simulations, evaluate the gradient uncertainty angle
        $\omega$ (see Figure~\ref{fig:angle_uncertainty}), and, if necessary,
        refine
        the estimates by running additional batches of $N$ simulations for both
        points.
        Because the stopping criterion depends on the pair, the same number of
        extra
        simulations is assigned to each point. This strategy could be further
        optimized
        by: i) distributing extra simulations according to each point’s
        individual
        uncertainty, and ii) reusing nominal-point simulations for multiple
        perturbation points.}
    \label{fig:workflow}
\end{figure}

In the following subsections, we give the remaining details. A graphical
representation of the workflow for a single nominal point is given in
Figure~\ref{fig:workflow}.

\subsection{Gradient Approximation}\label{Subs:gradient}
Since we are considering stochastic systems---with internal noise and
fluctuations---we, as outlined above, are interested in the sensitivity of the
expected value of the model observable,
$\mathbb{E}[f(\mathbf{x})]=\bar{f}(\mathbf{x})=\bar{y}$ to perturbations in
nominal points $\mathbf{x} \in X$. Since we are only interested in the
observable at a single simulation time point, we omit time from the notation.
We characterize the sensitivity via the estimation of the gradient vectors
\begin{equation}
    \nabla \bar{f}(\mathbf{x}) = \left[ \frac{\partial \bar{f}}{\partial
            x_1}(\mathbf{x}),\dots,\frac{\partial \bar{f}}{\partial
            x_n}(\mathbf{x})\right].
\end{equation}
To approximate the gradient vector, we employ a finite difference
approximation, specifically using forward
differences~\cite{turanyi1990sensitivity, burden2001, fu2015stochastic}. While
we acknowledge the variety of methodologies available for gradient estimation,
this study focuses on presenting the general framework rather than exploring
specific techniques. We consider the forward differences method to be
appropriate for its computational simplicity and ease of
interpretation~\cite{rabitz1983sensitivity}.

Given $\mathbf{x} = (x_1, \ldots, x_n) \in X$, we calculate perturbation
vectors by incrementing one dimension at a time while all other dimensions are
held unperturbed. That is, for each dimension $j \in \{1, \ldots, n\}$, we
create the perturbed point $\mathbf{x}'_j = (x_1, \ldots, x_j + h, \ldots,
    x_n)$, where $h > 0$ denotes a small perturbation. %This one-at-a-time perturbation ensures that the effect of each parameter on the observable is estimated independently of the others. 
As $h$ approaches zero, the forward
difference
\begin{equation}
    Z_{\mathbf{x},j}=\frac{\bar{f}(\mathbf{x}'_j)
        - \bar{f}(\mathbf{x})}{h}
\end{equation}
tends towards the partial derivative $\frac{\partial \bar{f}}{\partial
        x_j}(\mathbf{x})$ evaluated at $h=0$. However, in stochastic systems,
$h$ needs
to be balanced: it should be small enough to reduce the truncation error in the
forward difference approximation, which is caused by ignoring higher-order
terms in a Taylor series approximation, but large enough to avoid
simulation-related errors, such as rounding errors or noise in the stochastic
simulations~\cite{gunawan2005sensitivity}. The approximated gradient vector is
given by
\begin{equation}
    \nabla \bar{f}(\mathbf{x}) \approx \left( \frac{\bar{f}(\mathbf{x}'_1)
        - \bar{f}(\mathbf{x})}{h}, \ldots, \frac{\bar{f}(\mathbf{x}'_n)
        - \bar{f}(\mathbf{x})}{h} \right)=\left(Z_{\mathbf{x},1}, \dots,
    Z_{\mathbf{x},n}\right).
\end{equation}

The resulting gradients can be graphically represented in a vector field,
thereby visualizing the model's sensitivity to changes in the input parameters
across the parameter space~$X$.

\subsection{Simulation Stopping Criterion: Gradient Range of Angle Uncertainty}

Given $\mathbf{x}\in X$, the model observable $f(\mathbf{x})$ is a stochastic
variable where the uncertainty is linked to the internal stochastic effects
within the system. The variable distribution, population mean
$\mathbb{E}[f(\mathbf{x})]=\bar{f}(\mathbf{x})$, and population standard
deviation $\sigma$ are unknown. To approximate $\bar{f}(\mathbf{x})$ we need to
run simulations a sufficient number of times. As a key component of our
methodology, this sample size will be dynamically adjusted based on the
uncertainty present in the gradient approximation. Then the mean of the final
sample is used as the estimate of $\bar{f}(\mathbf{x})$. We will break this
down in the following.

\begin{figure}
    \centering
    % First subplot
    \begin{subfigure}[t]{0.3\textwidth}
        \centering
        \includegraphics[width=\textwidth]{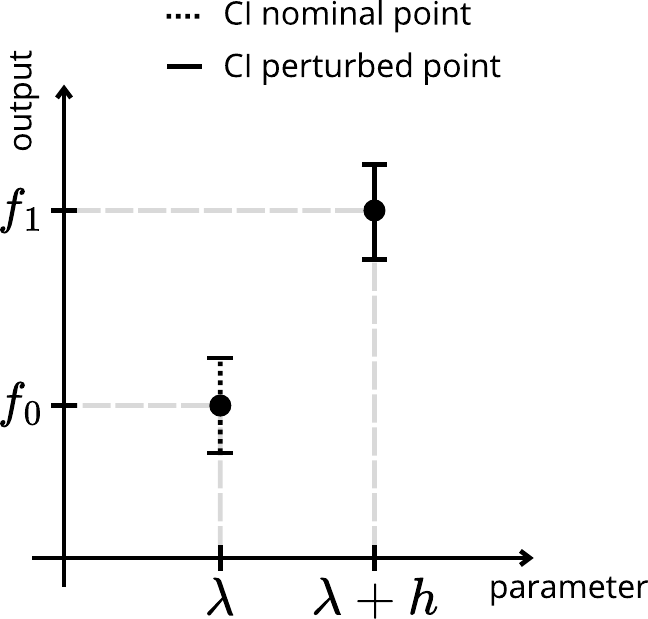}
        \caption{}
        \label{fig:angle_a}
    \end{subfigure}
    % Second subplot
    \begin{subfigure}[t]{0.3\textwidth}
        \centering
        \includegraphics[width=\textwidth]{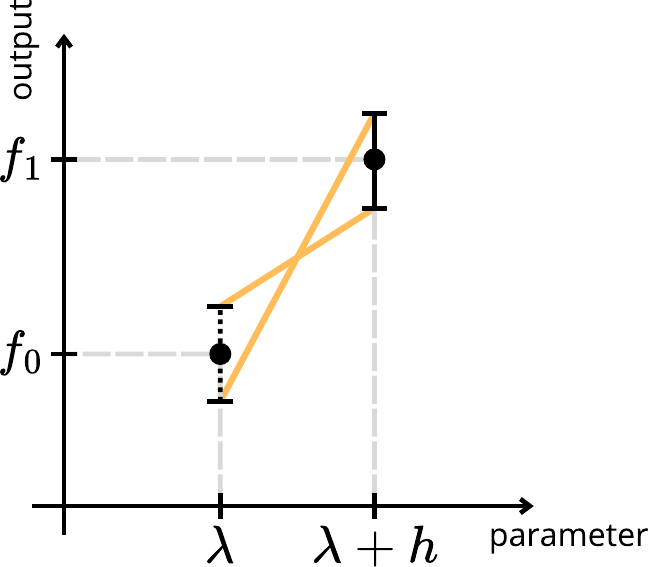}
        \caption{}
        \label{fig:angle_b}
    \end{subfigure}
    % % Third subplot
    % \begin{subfigure}[t]{0.24\textwidth}
    %     \centering
    %     \includegraphics[width=\textwidth]{figures/angles_figures/c.pdf}
    %     \caption{}
    %     \label{fig:angle_c}
    % \end{subfigure}
    % Fourth subplot
    \begin{subfigure}[t]{0.3\textwidth}
        \centering
        \includegraphics[width=\textwidth]{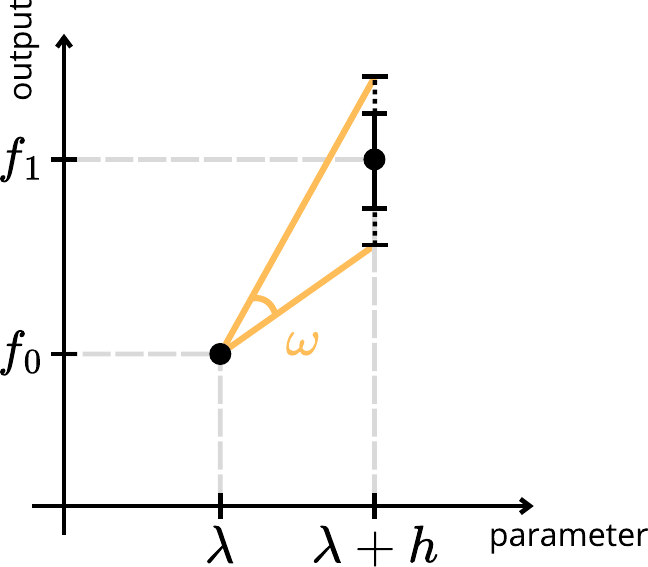}
        \caption{}
        \label{fig:angle_d}
    \end{subfigure}

    \caption{
        Angular range of gradient uncertainty $\omega$.
        \subref{fig:angle_a} The individual confidence intervals (CIs) for both
        the nominal point and its perturbation in one dimension.
        \subref{fig:angle_b} The extreme possible gradients.
        % \subref{fig:angle_c} Combination of the confidence intervals to         simplify the range of uncertainty.
        \subref{fig:angle_d} The angular range within which the possible
        gradients exist.
    }
    \label{fig:angle_uncertainty}
\end{figure}

To obtain a reliable gradient approximation,
our approach is to minimize the uncertainty in the slopes $Z_{\mathbf{x},1},
    \dots, Z_{\mathbf{x},n}$.
For a slope $Z_{\mathbf{x},j}$ this uncertainty can be regarded as an
\textit{angle} that represents the
range inside which many possible slopes can exist---each potentially being the
one connecting the
true population means of $f(\mathbf{x})$ and $f(\mathbf{x'_j})$.
When running $N$ initial simulations to obtain $f(\mathbf{x})$ and
$f(\mathbf{x'_j})$,
we can calculate the \SI{95}{\percent} confidence intervals (CI) for
$\bar{f}(\mathbf{x})$ and $\bar{f}(\mathbf{x'_j})$.
Recall that the characteristics of CIs is that with \SI{95}{\percent}
confidence, the true population mean lies somewhere within it (given suitable
assumptions on the probability distribution in question).
In other words, the interval defines a range of plausible values for the true
mean, based on the observed data. The width of the interval,
determined by the margin of error (MoE), represents the scope within which the
gradient likely lies. Geometrically, one could draw multiple slopes by connecting any value within the CI around $\bar{f}(\mathbf{x})$ to any value within the CI around $\bar{f}(\mathbf{x'_j})$ (\ref{fig:angle_a}). This yields a family of slopes, between the extreme possible gradients (\ref{fig:angle_b}). The same extreme gradients appear if we combine the two intervals into a single extended one (right side of figure \ref{fig:angle_d}), and choosing any slope connecting $\bar{f}(\mathbf{x})$ to this new extended interval---a valid candidate for the true gradient. This collection of slopes is captured by an angle
$\omega$, the range of gradient uncertainty.
Figure~\ref{fig:angle_uncertainty} illustrates $\omega$.
As the number of simulations increases to $2N, 3N,\dots$
(see~Figure~\ref{fig:workflow}), the confidence interval narrows, resulting in
a decrease in the uncertainty angle $\omega$. This angle is a way to quantify
and control the confidence in the gradient estimation and hence of the outcome
of the sensitivity analysis. In this paper, we set as our goal to run
simulations to achieve $\omega<\qty{5}{\degree}$ before timeout, but other
angle uncertainty thresholds could, of course, be chosen. We should note that this angle threshold should always be chosen in relation to the scaling (i.e., choice of units) of both the perturbed input dimension and the output axis. A similar comment applies to the relation between scales (units) of each of the input dimensions, which should be in meaningful proportions to each other. Such choices rely on domain knowledge for each use case. 
% Additionally, the partial
% derivatives can be scaled in order to make them independent of the
% units of $f$ and $x_j$~\cite{rabitz1983sensitivity, saltelli2005sensitivity}.
In the systems
analyzed here, all parameters correspond to rule rate constants with equal
numerical ranges and units, so no scaling was required.
% However, in cases where parameters have different magnitudes or units, an
% appropriate normalization step should be applied prior to the gradient
% computation, to reflect the sensitivity correctly.
%

This approach, where we dynamically adjust the sample size based on the
observed variability, not only gives control over the the precision of the
analysis, but also ensures that just enough data is collected per point to
achieve the desired precision, avoiding unnecessary sampling and saving time
and resources.

\subsection{Computing Confidence Intervals}

Our approach relies on statistical principles to ensure reliability in our
estimates. Given a sample $\{y_1, \ldots, y_{N}\}$, by the Law of Large
Numbers, the sample mean $\bar{y} = \frac{\sum_{i=1}^{N} y_{i}}{N}$ is an
unbiased estimator of the true population mean $\mu$, i.e., as $N$ increases,
$\bar{y}$ converges to $\mu$. By the Central Limit Theorem, for $N$ large
enough the distribution of the sample mean $\bar{y}$ approximates the normal
distribution with mean $\mu$ and standard deviation $\sigma/\sqrt{N}$,
regardless of the shape of the original distribution (assuming finite
variance)~\cite{bertsekas2008introduction, ross2017introductory,
    kwak2017central}. This implies that we can quantify the uncertainty in the
estimate $\bar{y}$ using confidence intervals.

Every time we run a new batch of simulations for a parameter vector
$\mathbf{x}$ (nominal or perturbed), we compute the sample mean $\bar{y}$ and
sample standard deviation $S$, given by
\begin{equation}
    S = \sqrt{\frac{\sum_{i=1}^{N} (y_i - \bar{y})^2}{N-1}}.
\end{equation}
For a confidence level $\alpha$, we then determine the two-sided $t$-value
$t_{N-1, \alpha/2}$ corresponding $N-1$ degrees of freedom. The margin of error
is then:
\begin{equation}
    \text{MoE} = t_{N-1, \alpha/2} \frac{S}{\sqrt{N}}.
\end{equation}
The confidence interval for the sample mean is then defined as:
\begin{equation}
    CI = \left[\bar{y}-\text{MoE}, \ \bar{y}+\text{MoE}  \right].
\end{equation}
This interval informs of the precision of the estimate $\bar{y}$ for the true
mean $\mu$. Wider intervals indicate more uncertainty due to great variability
in our data or smaller sample sizes. Therefore, the confidence interval serves
as a way to interpret the reliability of simulation outcomes.

\subsection{Global Sensitivity Coefficients}\label{subsec:SC}

To better understand the relative impact of each input variable on the expected
model observable $\bar{f}(\mathbf{x})$, we calculate a numerical sensitivity
coefficient for each input variable averaged across the set of sampled points in
the parameter space. For this
purpose, given the approximated gradient vector $\nabla \bar{f}(\mathbf{x})$ at
$\mathbf{x} \in X'$, we extract the absolute value of each component:
\begin{equation}
    \left| Z_{\mathbf{x},1} \right|, \ldots, \left| Z_{\mathbf{x},n} \right|.
\end{equation}
We then define the global\footnote{Global here refers to looking at input variables over their entire domain of definition and then aggregating the resulting information into single measures of parameter importance for the entire domain\cite{rabitz1983sensitivity}.} sensitivity coefficient for the $j$th input variable $x_j$,
$j
    \in \{1, \ldots, n\}$, as the average of these absolute values across all
sampled points~\cite{rabitz1983sensitivity}:
\begin{equation}
    SC_j = \frac{1}{M} \sum_{i=1}^{M} \left| Z_{\mathbf{x}^i,j}\right|,
\end{equation}
where $M$ denotes the total number of sampled points.

This measure represents the average absolute rate of change of
$\bar{f}(\mathbf{x})$ with respect to the $j$th variable across the sampled
parameter space. It provides a quantitative
assessment of the observable's sensitivity to a given variable based on the approximated gradients. 
%We note that these sensitivity
% coefficients have limitations. Sensitivities may vary over time if the model is
% dynamic, and our measure reflects only the chosen simulation time (or
% steady-state values if the simulation is run to equilibrium). Furthermore, the
% coefficients are averaged over a finite, bounded region of parameter space, so
% they do not capture behavior outside these ranges. Therefore, while useful for
% comparing the relative importance of variables, these coefficients should be
% interpreted with these considerations in mind.

%%%%%%%%%%%%%%%%%%%%%%%%%%%%%%%%%%%%%%%%%%%%%%%%%%%%%%%%%%%%%%%%%%%%%%%%%%%%%%%
\section{Application Examples}\label{Sec:Examples}
%%%%%%%%%%%%%%%%%%%%%%%%%%%%%%%%%%%%%%%%%%%%%%%%%%%%%%%%%%%%%%%%%%%%%%%%%%%%%%%

In this section we will show applications of our methodology.
The approach itself is general and can be applied in many domains---we here
choose chemical systems as an illustrative setting in which rule rate
constants constitute a natural choice of parameters to perform sensitivity
analysis on. We use the cheminformatics software
MØD~\cite{andersen2016software} as our vehicle for modeling the systems in a
rule-based fashion---in particular, we use MØD's stochastic simulation
module~\cite{machado2025rulebasedgillespiesimulationchemical} for performing
Gillespie-based stochastic simulations. In rule-based modeling, a single rule
represents several reactions with the same mechanism, hence systems can often
be modeled by a small number of reaction rules. Assigning reaction rate
constants to rules rather than individual reactions thus helps keeping the
number of parameters in the sensitivity analysis down.	In all our experiments,
the input parameters will be the rate constants for the rules of the
system, while custom observable measures will be defined for each system.

\begin{figure}
    \centering
    \includegraphics[width=0.35\linewidth]{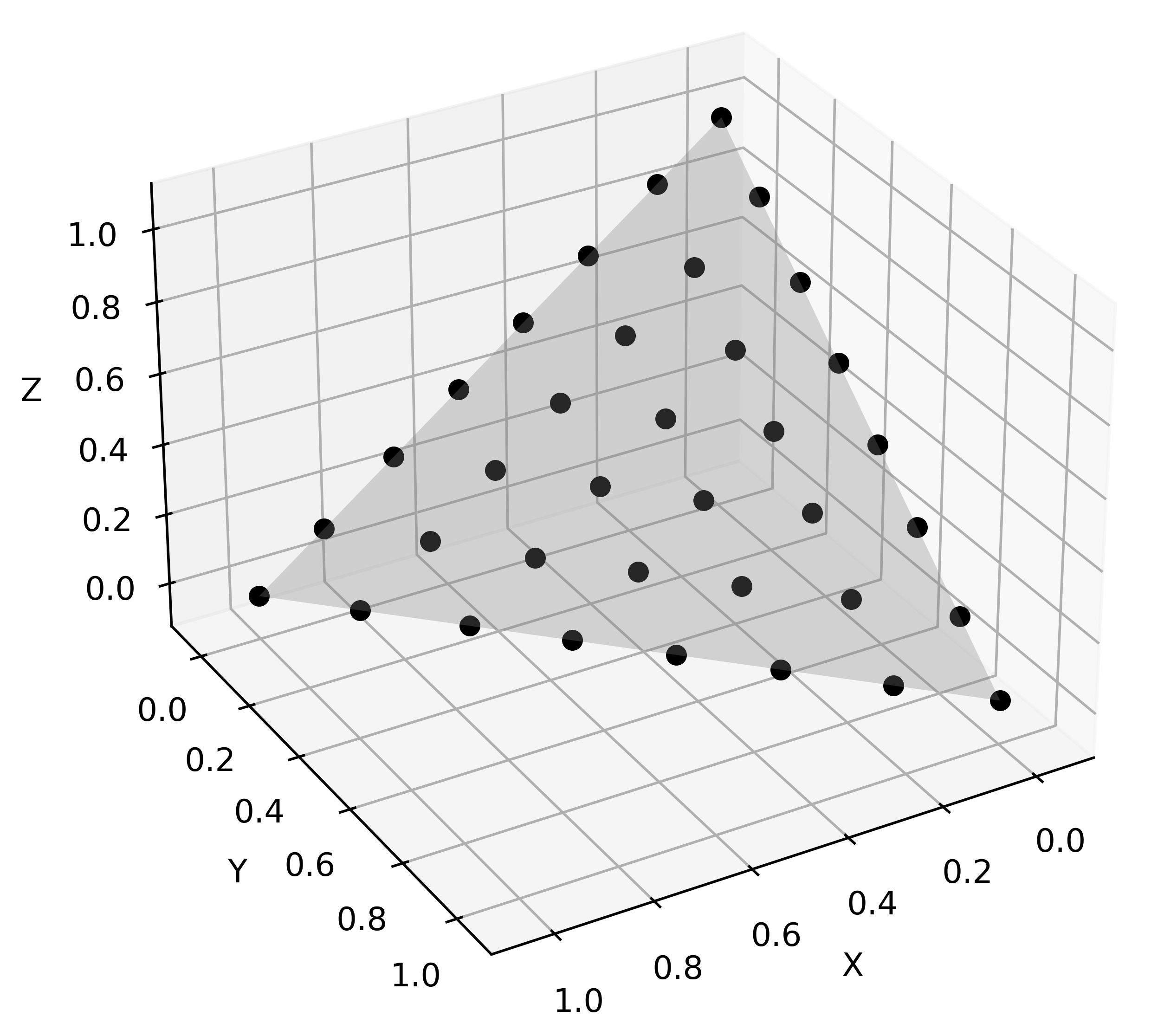}
    \caption{Input parameter points in the $2$-simplex.}
    \label{fig:2simplex}
\end{figure}

We will be working with three-dimensional parameter spaces $X \subseteq
    \mathbb{R}^3_{\geq 0}$. In particular, we consider points $\mathbf{x} =
    (x_1,
    x_2, x_3)$ in the $2$-simplex

\begin{equation}
    \mathbf{x} \in \left\{ (x_1, x_2, x_3) \in \mathbb{R}^{3} \,\middle|\,
    \sum_{i=1}^{3} x_i = 1, \, x_i \ge 0 \right\},
\end{equation}

as illustrated in Figure~\ref{fig:2simplex}.
For this example, scaling all rule rates by a common factor only changes the speed at which
the chemical system evolves, not the (probability distribution of the) outcome
of the simulation. Thus, for a given point in parameter space, there is no need to consider the rest of the points lying on the line starting at origin and passing through the point. Therefore, we normalize the rates such that their sum is a
constant, here chosen to be one. In Gillespie simulation, this sum determines the speed with which the system develops, and we here ensure that this is the same for all the nominal points considered. 

We present the main results of the sensitivity analysis as a series of 2D
plots. Our data (points on the simplex and their estimated gradient vector) are
of course 3D objects, but such data is notoriously hard to present in a clearly
readable fashion, hence this choice of presentation. To help the
interpretation, we show the points and their gradients in two different types
of projections: a projection onto the plane of the simplex and three
projections onto the three basic 2D planes of the 3D coordinate system (as an
example, see Figure~\ref{fig:product-combined} below).
In the latter type of projections, the two axes represent the rate constants of
two of the three reactions (the remaining rate constant is given by the
requirement that the three rates sum to one). The vectors shown are the
gradient at each point, subjected to the same projection and colored to show
the value of the observable at the point in question. To improve readability,
the vectors have been scaled so that the maximum length appearing equals the
distance between points.

We consider two concepts that will help analyze our results---both
encapsulating a notion of robustness, but in different ways.
One concept is \textit{local sensitivity}, which refers to how much the average
value of the observable changes in response to small perturbations of the input
parameters in a given region of the input space.
A region is considered robust in this sense (low local sensitivity) if such
perturbations lead to relatively minor changes in the observable values.
Another concept is \textit{gradient reliability}, which refers to the
consistency of gradient estimates in light of the stochastic nature of the
simulations.
A result is considered reliable if repeated simulations under the same
conditions yield similar outcomes, represented in our case by smaller
confidence intervals (or equivalently, a smaller range of angle uncertainty).
See Figure~\ref{fig:angle_d}.

As will be shown in the examples, we can encounter situations where the
observable displays little relative change to perturbations of the input
parameters (low local sensitivity), but where the uncertainty intervals for
gradients are large due to internal stochastic noise (low gradient
reliability). The other way around, a gradient might be produced by consistent
estimates (high reliability), even if small perturbations cause large
observable changes (high local sensitivity). In fact, all four combinations of
sensitivity and reliability are possible.

\subsection{Michaelis-Menten Kinetics}

\subsubsection{Description}

The Michaelis-Menten
kinetics~\cite{schuster2016stochasticity,michaelis1913kinetics}
model general enzyme catalysis.
Here, we consider the simplified reaction scheme, consisting of the reactions
\begin{equation}
    \begin{tikzcd}
        S + E \arrow[r, shift left, "\text{binding}"] & C \arrow[l, shift left,
            "\text{unbinding}"] \arrow[r, "\text{catalysis}"] & E + P,
    \end{tikzcd}
\end{equation}
where $S$ represents the substrate, $E$ is the enzyme, $C$ is the complex
resulting from binding the substrate to the enzyme, and $P$ is the product.

\begin{figure}
    \centering
    \includegraphics[width=0.48\linewidth]{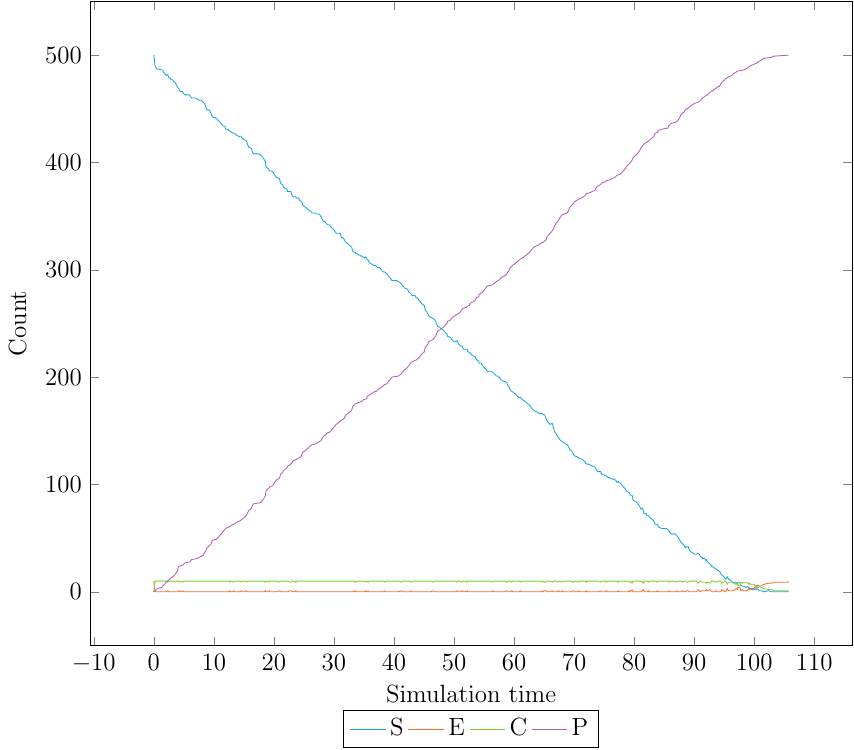}
    \caption{Stochastic simulation of the evolution (simulation time units) of
        Michaelis-Menten dynamics.}
    \label{fig:evol-michaelis}
\end{figure}

We apply the sensitivity analysis procedure described above and consider the
reaction rate constant of each of the three reactions as the input parameters.
For observable, we choose the simulation time (by which we mean units of time
in the simulated process) until all substrate $S$ is consumed. The initial
molecular counts are $500$ units of substrate $S$ and $10$ units of enzyme $E$.
As an illustrative example, one simulation of the evolution of the
Michaelis-Menten dynamics using the stochastic simulation module in MØD
produces the output in Figure~\ref{fig:evol-michaelis} when all rate constants
are~$0.5$.
For the sensitivity analysis, both the perturbation size $h$ and the initial number of simulations (and batch size) $N$ were chosen based on preliminary tests. We set $h$ to the smallest perturbation that produced stable gradient directions across repeated runs, ensuring a reliable finite-difference approximation. The batch size $N=50$ offered a reasonable baseline per nominal-perturbation pair and was used as the increment for adaptive refinement, with additional batches added only when higher precision was needed. Further experimental design choices are shown below.
\begin{itemize}
    \item Initial number of simulations per input point $\mathbf{x}\in X$:
          $N=50$.
    \item Runtime limit for simulating each $\mathbf{x}\in X$: $300$ seconds
          wall clock time.
    \item Confidence level for margin of error calculation: $\alpha = 0.95$.
    \item Target gradient-uncertainty angle: $\omega=\qty{5}{\degree}$.
    \item Size of the perturbation for gradient estimation: $h = 0.071$ (\qty{50}{\percent} of the distance between input points).
\end{itemize}

\subsubsection{Sensitivity Analysis}

\begin{figure}
    \centering
    \begin{subfigure}[b]{0.50\textwidth}
        \centering
        \hspace*{3.75em}%

        \includegraphics[scale=0.5]{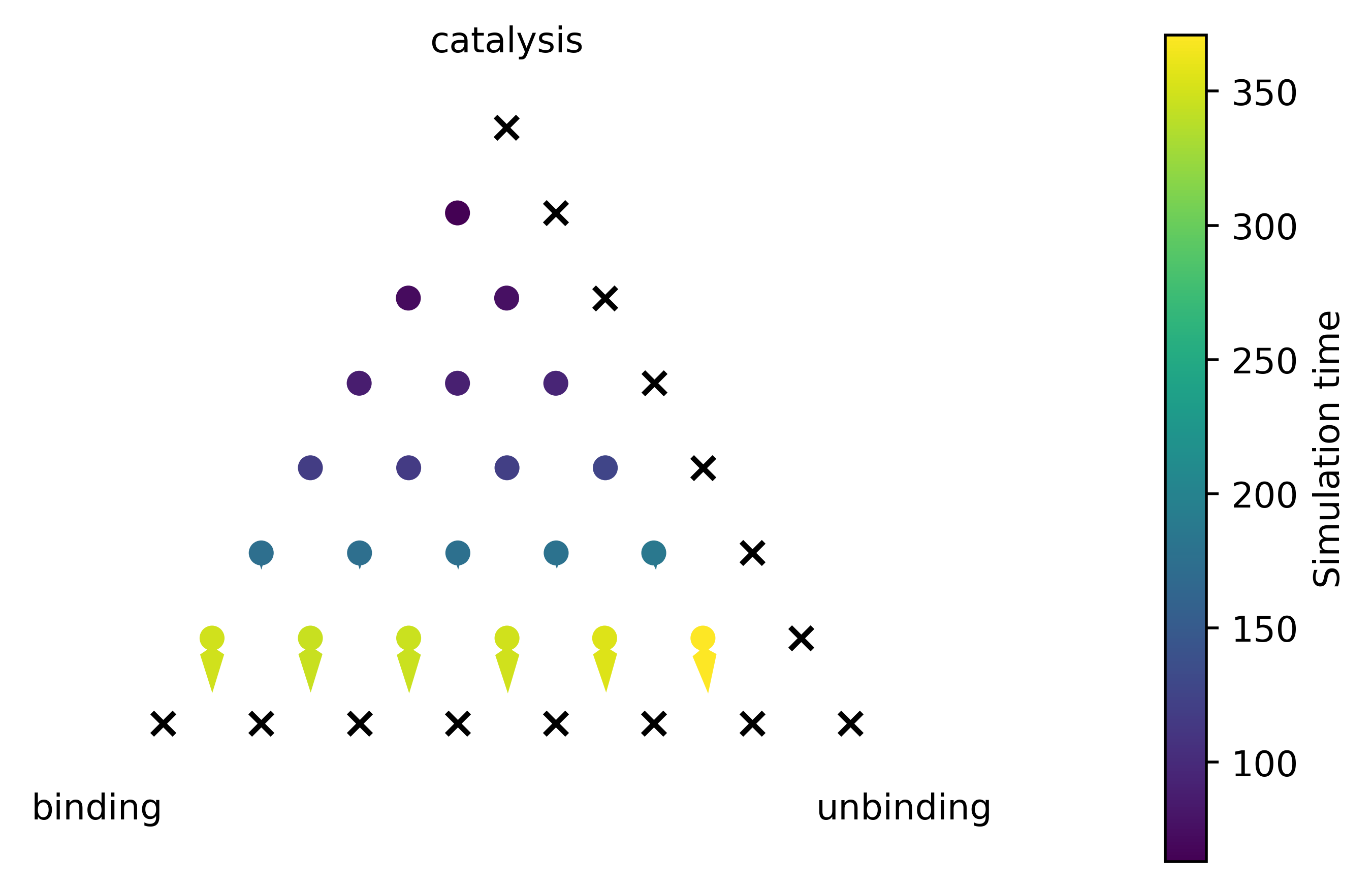}
        \caption{Projected gradient vectors.}
        \label{fig:fig1}
    \end{subfigure}%
    \begin{subfigure}[b]{0.50\textwidth}
        \centering
        \hspace*{3.75em}%

        \includegraphics[scale=0.5]{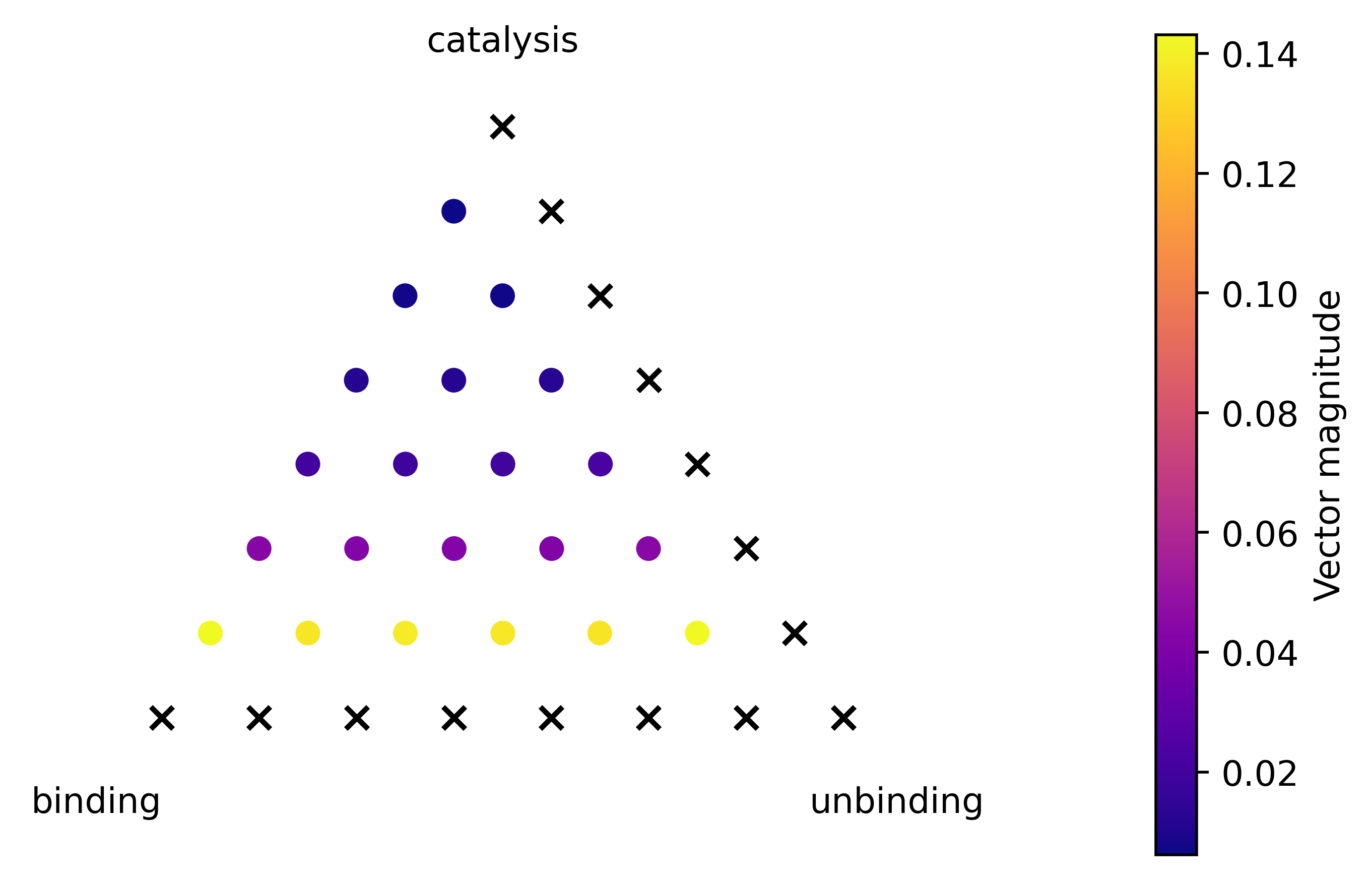}
        \caption{3D magnitude colour map.}
        \label{fig:fig2}
    \end{subfigure}

    \begin{subfigure}[b]{\textwidth}

        \includegraphics[width=\linewidth]{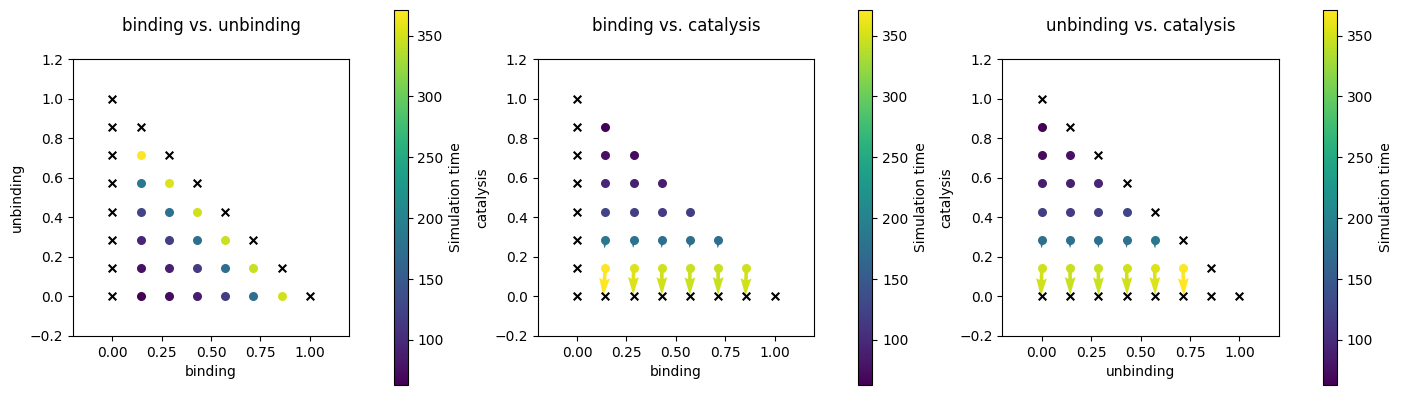}
        \caption{Gradient vectors projected into each of the 3 basic 2D planes
            of the 3D coordinate system.}
        \label{fig:product}
    \end{subfigure}
    \caption{Sensitivity analysis of the Michaelis-Menten Kinetics.
        \subref{fig:fig1} Gradient vectors projected onto the 2-simplex plane.
        Each point represents a sampled parameter set; the arrow at each point
        shows
        the local sensitivity (gradient vector) of the observable (simulation time to consume $S$) with respect
        to all
        three rate constants. The color of each point and vector indicates the
        value of
        the observable. Points with very small
        or zero
        gradient magnitude appear without a visible arrow.
        \subref{fig:fig2} Each sampled point colored according to the magnitude
        of the full, non-projected gradient vector; no arrows are shown in this
        panel.
        This highlights regions of high and low sensitivity.
        \subref{fig:product} Gradient vectors projected onto the three basic 2D
        planes of the 3D coordinate system, with colors indicating the
        observable as in
        panel (a). From these plots, we observe that sensitivity is largest
        with
        respect to the catalysis rate constant, especially when it is close to
        zero,
        whereas binding and unbinding rate constants contribute little to the
        variation
        in the observable.
    }

    \label{fig:product-combined}
\end{figure}

\paragraph{Local Analysis Across Parameter Space.}
Figure~\ref{fig:product-combined} depicts our results. These show that the
system is most sensitive to changes in the catalysis rate constant, especially
for values of it close to zero (but non-zero), as this is where vector
magnitudes are largest.
In contrast, variations in binding and unbinding rate constants lead to little
change in the observable, indicating low relative sensitivity. In panels
(\subref{fig:fig1}) and (\subref{fig:product}), the points and arrows are
colored according to the value of the observable (simulation time to consume
$S$), while in panel (\subref{fig:fig2}), points are colored according to the
magnitude of the full, non-projected gradient vector, highlighting regions of
high and low sensitivity; no arrows are shown in this panel. Arrows in
(\subref{fig:fig1}) and (\subref{fig:product}) indicate both the direction and
relative
magnitude of the local sensitivity; points with very small or zero gradient
magnitude appear without a visible arrow. Points marked by
a cross indicate that no observable was produced during the simulation: these
are the points where either the binding or the catalysis reactions had a rate
constant of zero, which prevents the substrate $S$ from being depleted.
Overall, the analysis shows that the catalysis rate constant is the key driver
of variation of this observable (the simulation time until $S$ is fully
consumed), and that there are clear parameter regions, aligned with the
magnitude of the catalysis rate constant, that minimize or maximize the value
of the observable.

\paragraph{Global Sensitivity Coefficients.}

In order to calculate the global sensitivity coefficients described in section
\nameref{subsec:SC}, we repeat the full experiment described in the bullet
points above five times and collect the data in Table~\ref{tab:sens_coeff_mm}.
This data summarizes the effect we observe in
Figure~\ref{fig:product-combined}, highlighting the strong relative sensitivity
of the observable measure (simulation time to consume $S$) to the catalytic rate constant.
We observe that the sensitivity coefficients show little variation across
experiments, as indicated by their small standard deviation.\\
\begin{table}
    \centering
    \sisetup{table-format=3.3}
    \begin{tabular}{@{}lSSSS@{}}
        \toprule
        Dimension & {Avg}  & {Min}   & {Max}   & {Std Dev} \\
        \midrule
        Binding   & 15.783  & 15.139  & 17.110  & 0.864     \\
        Unbinding & 10.001  & 8.292   & 12.330  & 1.747     \\
        Catalysis & 382.688 & 378.935 & 385.929 & 2.814     \\
        \bottomrule
    \end{tabular}
    \caption{Statistics for the sensitivity coefficient for each input
        parameter.}
    \label{tab:sens_coeff_mm}
\end{table}
\\
\noindent\textit{Conclusion of the sensitivity analysis:} The (simulation) time until all substrate $S$ is consumed is most sensitive to changes in the catalysis rule rate constant, especially when this parameter takes values close to $0$.

\subsubsection{Uncertainty Analysis}

\begin{figure}[h]
    \centering
    \begin{subfigure}[t]{0.32\textwidth}
        \centering
        \includegraphics[width=\textwidth]{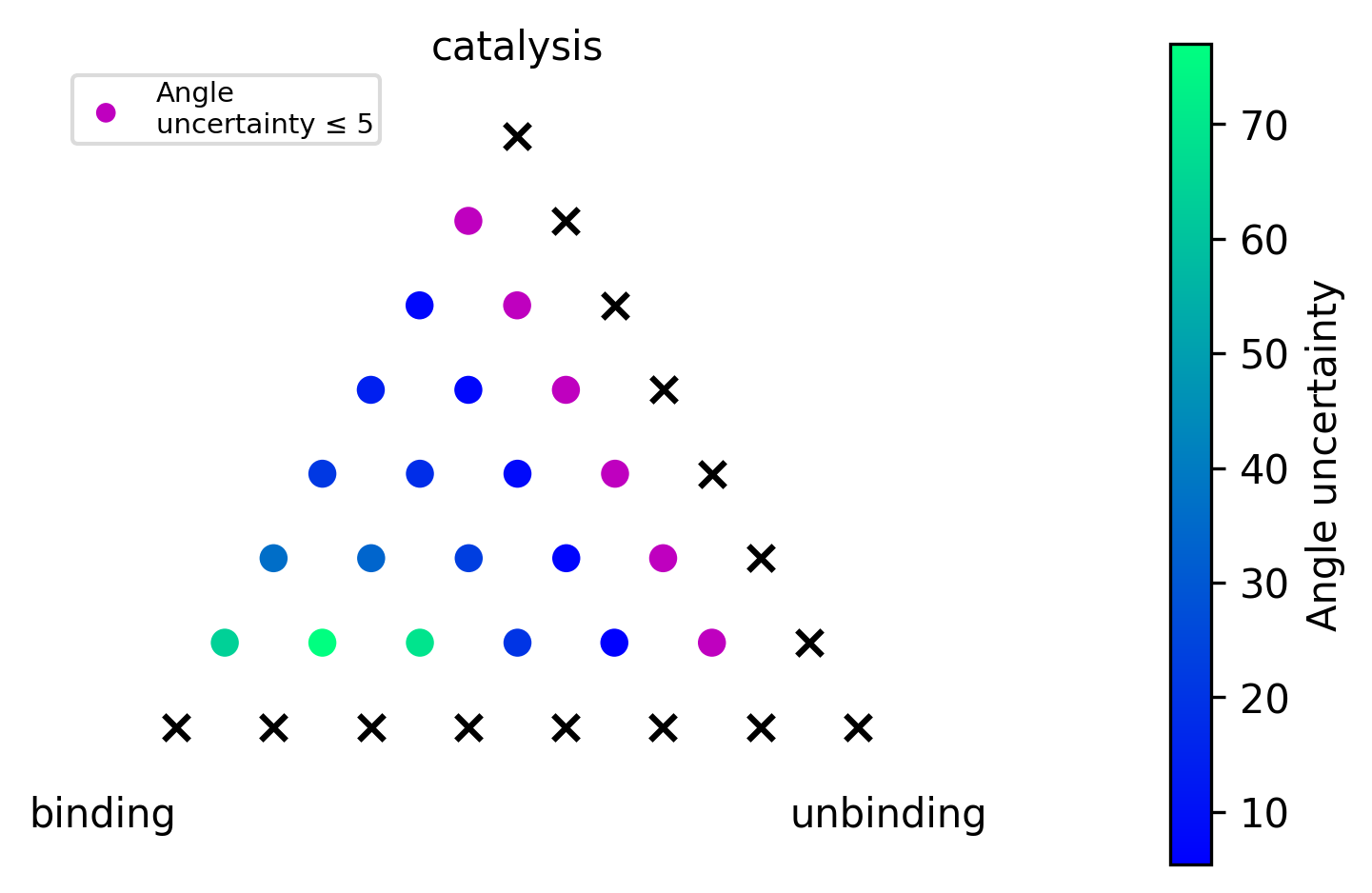}
        \caption{Binding.}
        \label{fig:uncertainty_binding}
    \end{subfigure}
    \hfill
    \begin{subfigure}[t]{0.32\textwidth}
        \centering
        \includegraphics[width=\textwidth]{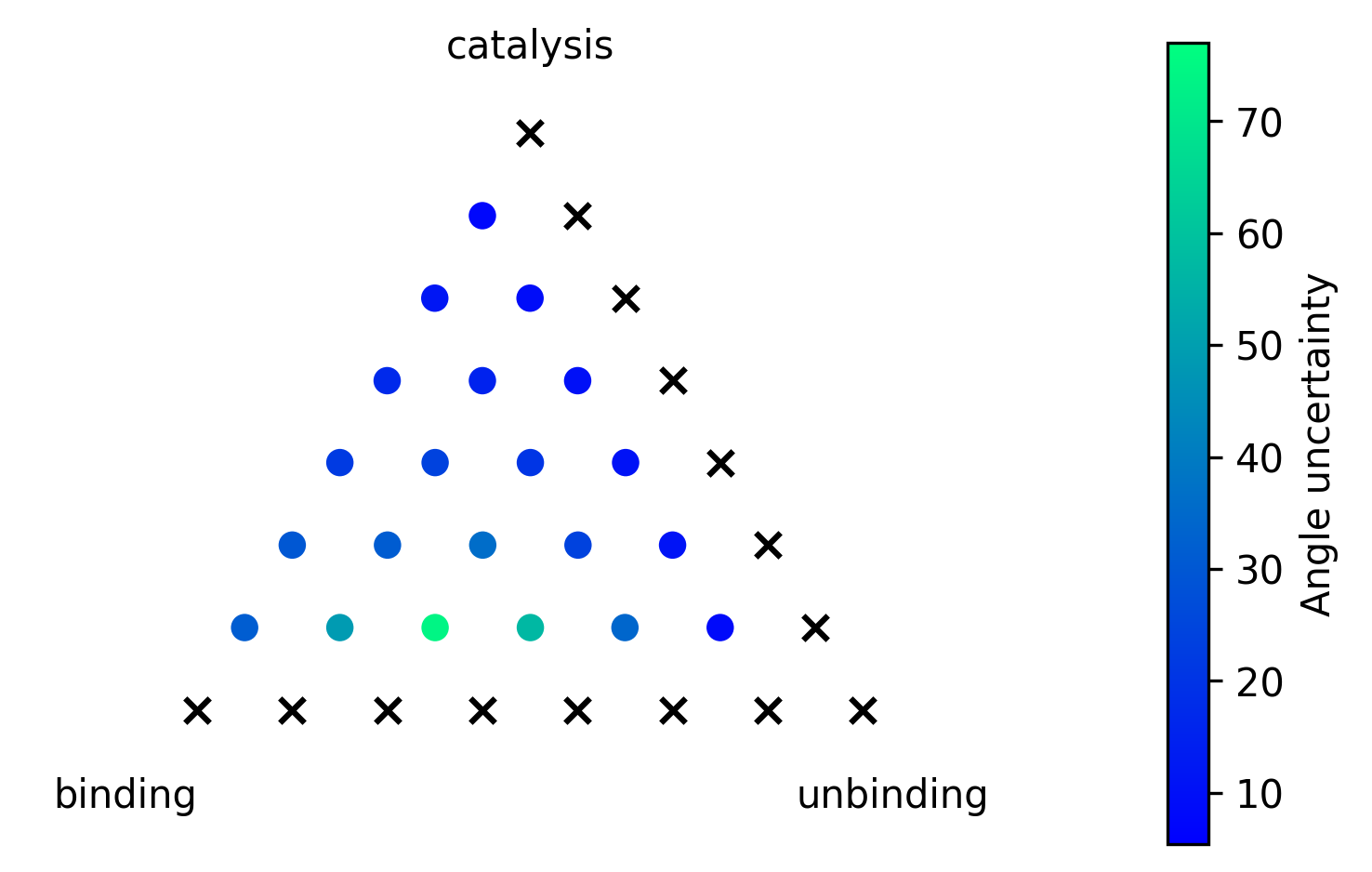}
        \caption{Unbinding.}
        \label{fig:uncertainty_unbinding}
    \end{subfigure}
    \hfill
    \begin{subfigure}[t]{0.29\textwidth}
        \centering
        \includegraphics[width=\textwidth]{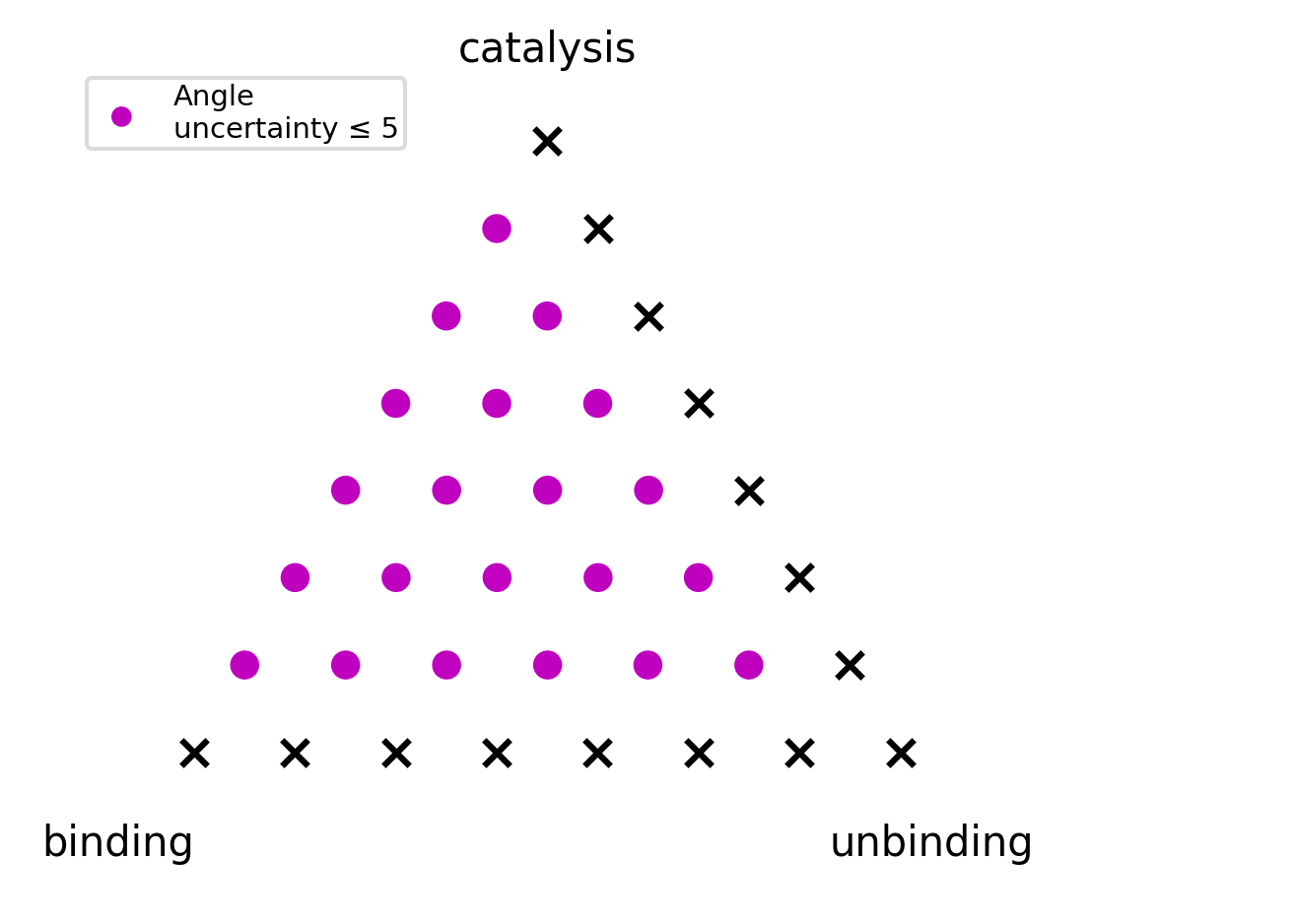}
        \caption{Catalysis.}
        \label{fig:uncertainty_catalysis}
    \end{subfigure}

    \caption{
      Gradient uncertainty across the parameter space. %Each subplot shows the corresponding angle uncertainty at each point, for each perturbed dimension.
      Each subplot considers one perturbed input dimension (i.e., one component of the gradient vector) and shows at each point in parameter space the angle uncertainty of this dimension.
      The average gradient-uncertainty angle across all points is for each dimension as follows:
    \subref{fig:uncertainty_binding} binding (20.905), 
    \subref{fig:uncertainty_unbinding} unbinding (25.364), 
    \subref{fig:uncertainty_catalysis} catalysis (0.774).
    }
    \label{fig:mm_uncertainty}
\end{figure}

Our initial objective was to reduce the gradient-uncertainty angle to at
most \qty{5}{\degree}, as a criterion for the number of simulations required.
However, the summary statistics in Table~\ref{tab:summary_mm} show that in the
available time, we achieved an average angle of approximately
\qty{15.7}{\degree}, with a substantial standard deviation of
\qty{19.6}{\degree} and a wide range from \qty{0.03}{\degree} to
\qty{77}{\degree}.
This spread highlights variability in the estimated gradient across
the parameter space and perturbed input dimensions, with some regions exhibiting more reliable estimates than
others, as shown in Figure \ref{fig:mm_uncertainty}. Regions with high angle uncertainty indicate where the local gradient
direction is less consistent, which may suggest that additional simulations in
these areas could help improve the reliability of the gradient estimates.
\begin{table}
    \centering
    \sisetup{table-format=2.3}
    \begin{tabular}{@{}lSS[table-format=1.3]SS@{}}
        \toprule
                                      & {Avg} & {Min} & {Max}  & {Std Dev} \\
        \midrule
        Gradient-Uncertainty Angle & 15.681 & 0.029 & 76.957 & 19.574    \\
        \bottomrule
    \end{tabular}
    \caption{Summary statistics for the gradient-uncertainty angle (across all nominal points and all three input dimensions), measured
        in
        degrees. Values show considerable variation across the input space.}
    \label{tab:summary_mm}
\end{table}

The catalysis is not just the parameter with the highest sensitivity, as seen in Figure~\ref{fig:product}, it also yields the most stable gradient estimates, as
reflected by its low gradient-uncertainty angle. This
indicates that when perturbing along the catalysis dimension, the gradient estimates
are relatively consistent across sampled points. In
contrast, the binding and unbinding dimensions show low sensitivity but high
angle of uncertainty, meaning that the range of possible gradient directions is
larger when perturbing along these dimensions. This is likely because the observable
changes very little in response to variations along these dimensions, so the local
gradient directions are less well-defined and exhibit higher angle uncertainty.
\\

% In other words, the angle uncertainty quantifies the variability of gradient
% estimates along a given perturbation direction. Low values indicate that
% perturbations along that dimension produce relatively consistent gradient
% directions, while high values indicate that the gradient direction can vary
% widely. This helps identify which directions are reliably estimated and which
% are more susceptible to noise.
% 

\noindent\textit{Conclusion of the uncertainty analysis:} The catalysis parameter yields the most reliable gradient estimates, consistently across the parameter space.

\paragraph{Extended Uncertainty Analysis: Different Runtime Limits.}

\begin{figure}
    \centering

    \includegraphics[width=0.6\linewidth]{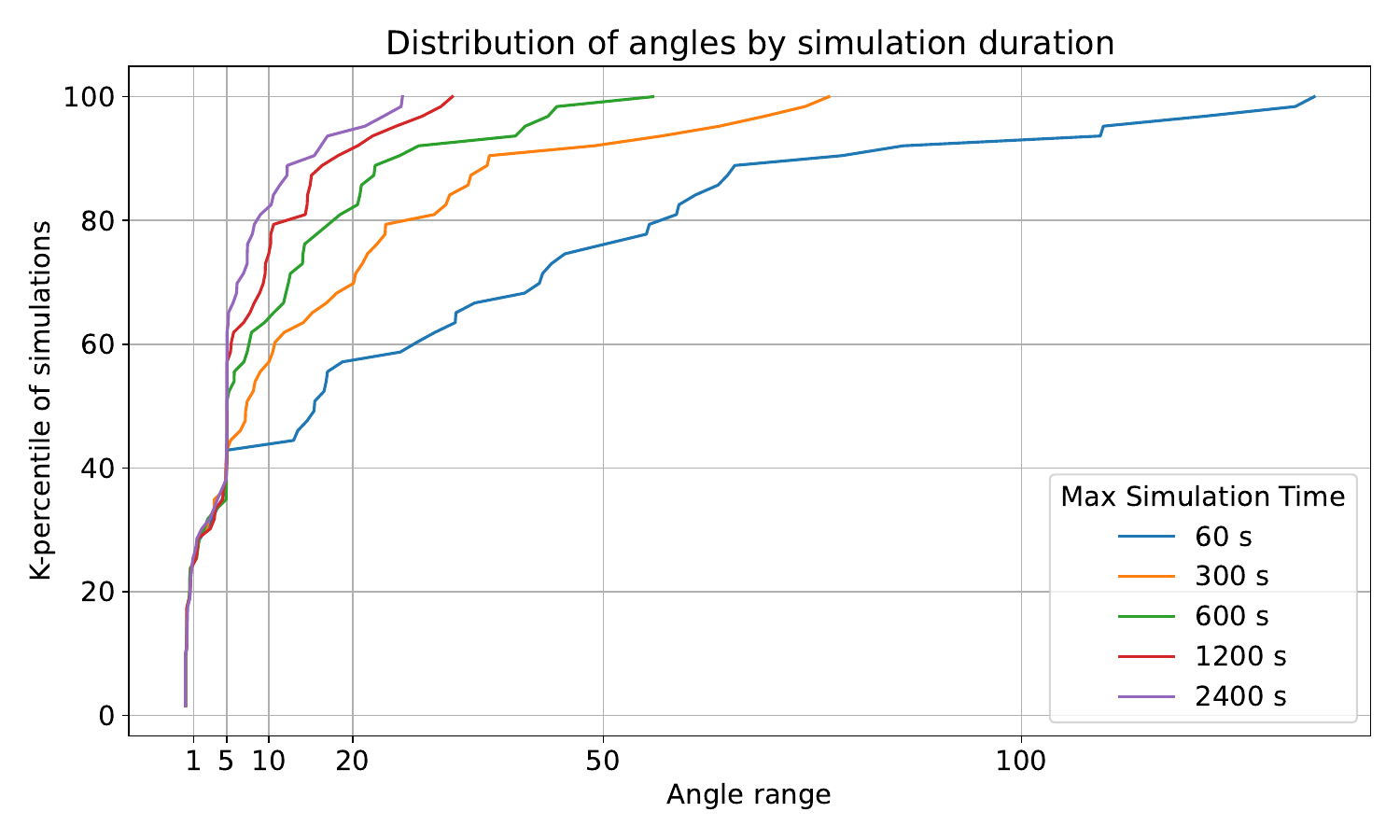}
    \caption{Percentile distribution of the number of times a given angle bound
        was reached, across all nominal points and perturbations. Each colored
        line
        corresponds to a different runtime limit used when simulating each
        nominal–perturbation pair.}
    \label{fig:mm-angle}
\end{figure}

Figure~\ref{fig:mm-angle} extends the above statistics on angles by plotting
the number of times (in percent across all nominal points and perturbation
experiments) that we achieved a given angle bound when we conducted the
experiments with different runtime limits. For instance, for a runtime limit of $300$ seconds wall clock time per nominal point and perturbation pair, %(which is the time used for the experiments above), 
% Notice that the results above corresponds to a runtime limit of $300$ seconds wall clock time per nominal point and perturbation pair (orange in Figure~\ref{fig:mm-angle}). For this specific value, 
about
\qty{43}{\percent} of the nominal-perturbation pairs achieved an angle smaller
than \qty{5}{\degree}, while around \qty{63}{\percent} achieved an angle
smaller than \qty{15}{\degree}. For a runtime limit of $1200$ seconds,
\qty{57}{\percent} of the nominal point and perturbation pairs achieved an
angle smaller than \qty{5}{\degree}. One noticeable difference is in the
maximum angle obtained for each time limit, ranging approximately from
\qty{26}{\degree} (purple) to \qty{135}{\degree} (blue). In larger percentiles,
we can observe more potential for better angle ranges when allowing longer
runtime limits.

\subsection{Formose Chemistry}\label{Subsec:Formose}

\subsubsection{Description}
We consider a simplified, rule-based model of the formose chemistry following
Andersen et al.~\cite{andersen2014generic}. Starting from formaldehyde and
glycolaldehyde, a small set of reaction rules—aldol addition (forward/backward)
and keto–enol tautomerization—generates the reaction network on-the-fly. We
simulate the dynamics using Gillespie’s stochastic simulation algorithm as
implemented in MØD. Our objective here is to analyze how the diversity of the
chemical space depends on these reaction-type rate constants.

The details of the experiment are as follows.
Consistent with our motivation to keep the parameter space tractable, we assign
rate parameters based on reaction types. Thus, the input parameters are the
rate constants of the reaction rules depicted in
Figure~\ref{fig:formose_rules}.
We also equate the rate constants of the two keto-enol rules (forward and
backward) such that these constitute a single input parameter.
The observable, aimed at measuring the diversity of the chemical space, is the
total number of different molecular species created throughout the simulation
after a fixed simulation time.
The initial state is composed of $500$ copies of formaldehyde and $50$ copies
of glycolaldehyde.
To limit the combinatorial explosion of compounds, the molecular size is capped
at eight carbon atoms.
Additionally, unlike in the previous example, here we utilize a simulation time
limit parameter in MØD for each simulation instance. Further experimental
design choices are shown below. Notice, like in the previous example, that both $N$ and $h$ where chosen empirically after preliminary tests with several values. 
\begin{itemize}
    \item Initial number of simulations per input point $\mathbf{x}\in X$:
          $N=50$.
    \item Runtime limit for simulating each $\mathbf{x}\in X$: $300$ seconds
          wall clock time.
    \item Simulation time limit of individual simulation instances: $20$
          simulation time units.
    \item Confidence level for margin of error calculation: $\alpha = 0.95$.
    \item Target gradient-uncertainty angle: $\omega=\qty{5}{\degree}$.
    \item Size of the perturbation for gradient estimation: $h = 0.071$ (\qty{50}{\percent} of the distance between input points).
\end{itemize}
\begin{figure}
    \centering
    \input{figures/formose_experiments/rules_figure_paper/inc}
    \caption{Reaction rules for the formose chemistry model. 
    	The keto-enol rules in \subref{rule: keto-enol F} and \subref{rule: keto-enol B} are inverses of each other.
    	This is similarly the case for the aldol addition rule in \subref{rule: aldol add F} and \subref{rule: aldol add B}.
	}
    \label{fig:formose_rules}
\end{figure}

We note that in systems like the formose chemistry, ODE modeling would not be
suitable for conducting this experiment since the reaction network is not known
in advance.
Here, we begin only with a set of initial compounds and a few reaction rules,
and the chemical space is then expanded on-the-fly by MØD.
In this context, sensitivity analysis must rely on averaging multiple
simulation realizations, given that each run may explore different reaction
paths.
This stochastic approach allows us to estimate the system's overall response to
changes in input parameters, without requiring a complete, static reaction
network---which in the case of formose chemistry can be of substantial
size---available beforehand.

\subsubsection{Sensitivity Analysis}

\begin{figure}
    \centering
    \begin{subfigure}[b]{0.50\textwidth}
        \centering
        \hspace*{3.65em}%

        \includegraphics[scale=0.5]{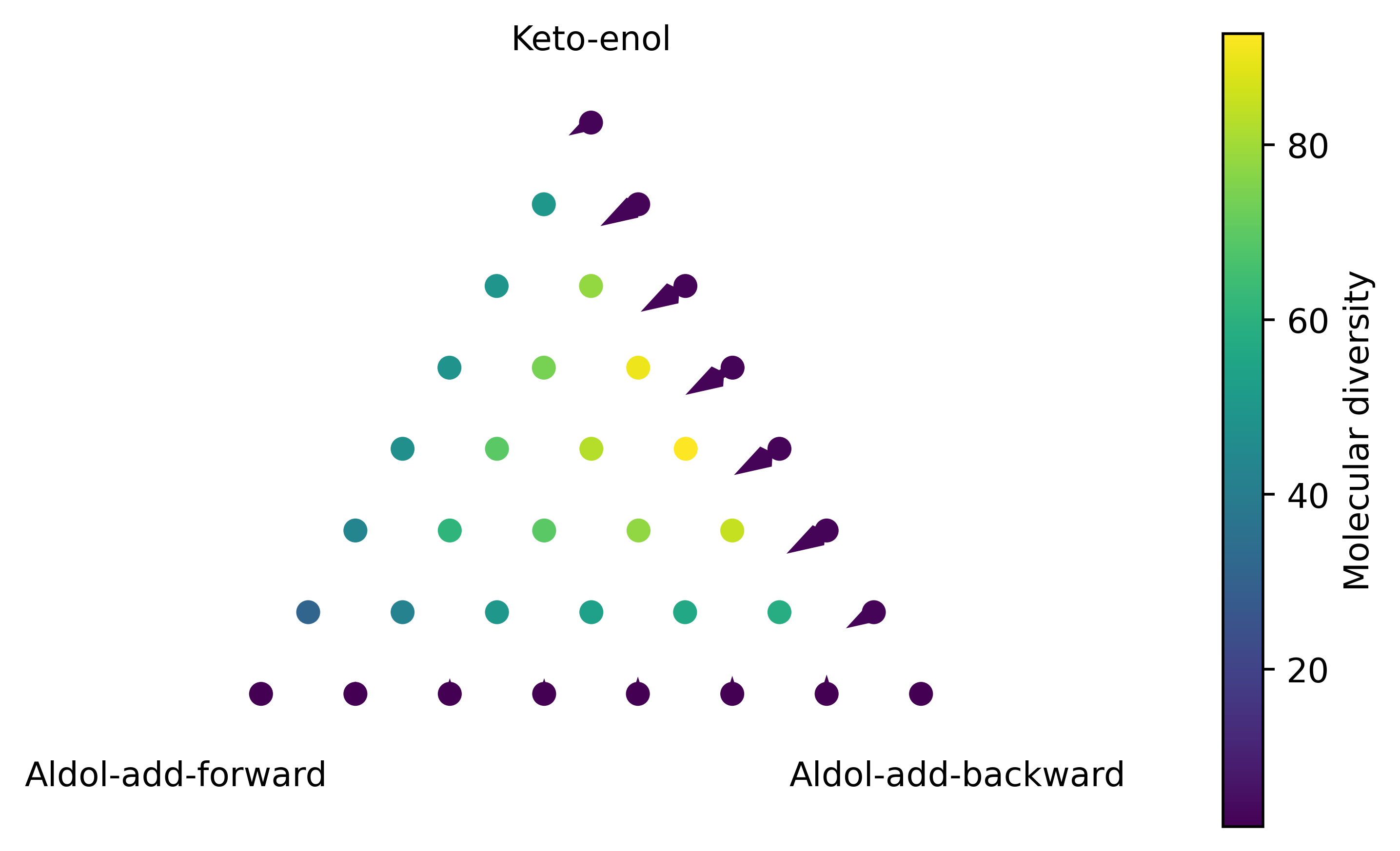}
        \caption{Projected gradient vectors.}
        \label{fig:fig1_formose}
    \end{subfigure}%
    \begin{subfigure}[b]{0.50\textwidth}
        \centering
        \hspace*{3.65em}%

        \includegraphics[scale=0.5]{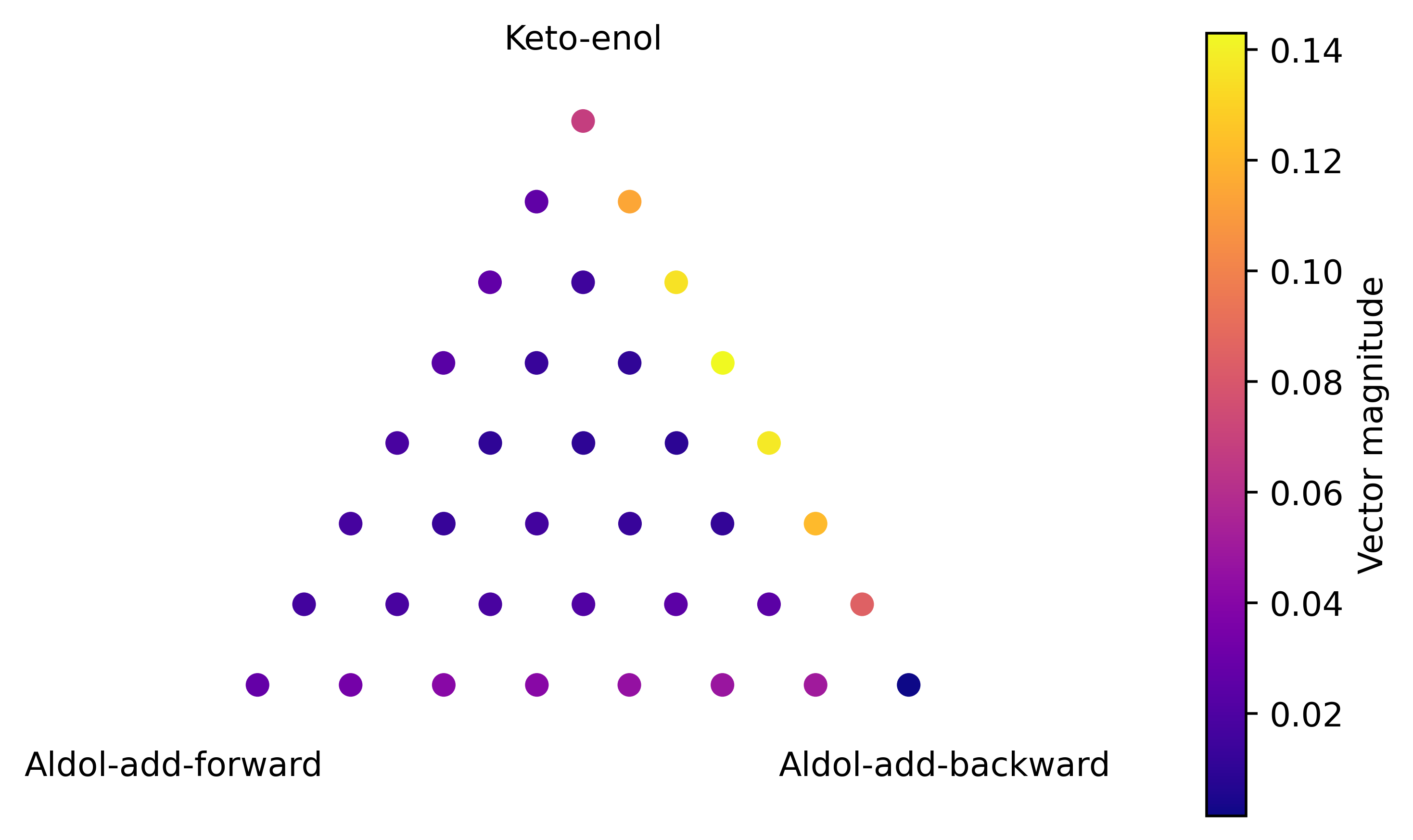}
        \caption{3D magnitude colour map.}
        \label{fig:fig2_formose}
    \end{subfigure}

    \begin{subfigure}[b]{\textwidth}

        \includegraphics[width=\linewidth]{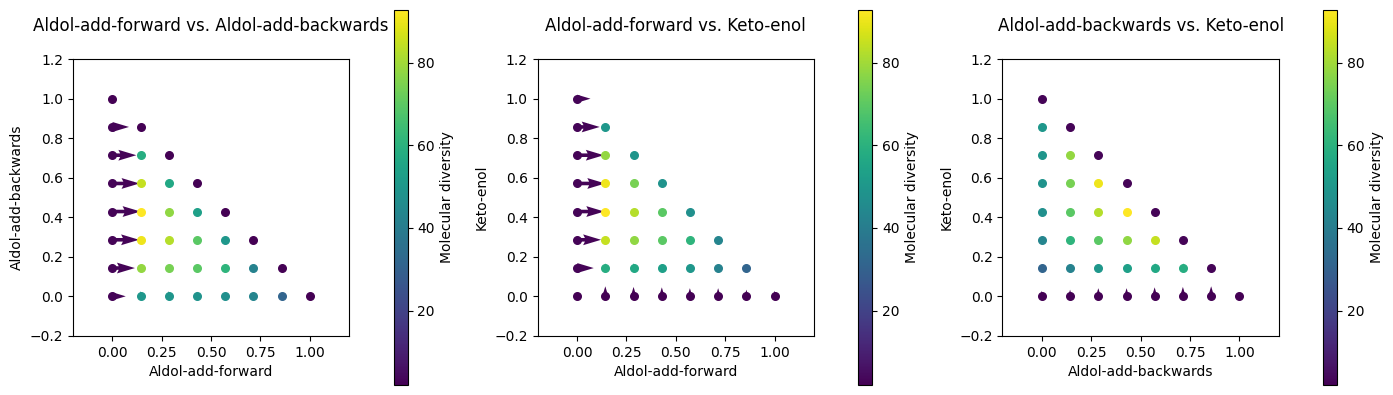}
        \caption{Gradient vectors projected into each of the 3 basic 2D planes
            of the 3D coordinate system.}
        \label{fig:product_formose}
    \end{subfigure}
    \caption{Sensitivity analysis of the formose model.
        \subref{fig:fig1_formose} Gradient vectors projected onto the 2-simplex
        plane.
        Each point represents a sampled parameter set; the arrow at each point
        shows
        the local sensitivity (gradient vector) of the observable with respect
        to all
        three rate constants. The color of each point and vector indicates the
        value of
        the observable (number of different molecular species created
        throughout the simulation). Points with very small
        or zero
        gradient magnitude appear without a visible arrow.
        \subref{fig:fig2_formose} Each sampled point colored according to the
        magnitude
        of the full, non-projected gradient vector; no arrows are shown in this
        panel.
        This highlights regions of high and low sensitivity.
        \subref{fig:product_formose} Gradient vectors projected onto the three
        basic 2D
        planes of the 3D coordinate system, with colors indicating the
        observable as in
        panel (a). These plots indicate that the observable, the number of
        different
        molecular
        species created throughout the simulation, is more sensitive to changes
        along
        the aldol-addition$\rightarrow$ dimension. The highest sensitivity is
        achieved in
        the region where this parameter equals zero.}
    \label{fig:product-combined_formose}
\end{figure}

\paragraph{Local Analysis Across Parameter Space.}
The results of the experiment are illustrated in
Figure~\ref{fig:product-combined_formose}.
We can observe that the system is most sensitive to changes along the
aldol-addition$\rightarrow$ dimension around the area where this parameter is zero.
This indicates that going from no activity to a little bit increases
dramatically the diversity of species in the chemical space (seen by the abrupt
change from purple to yellow points).
The relatively high sensitivity of the observable to changes along this dimension in
this specific region of the space creates the visual effect that changes in
other dimensions, or in other regions, do not have an impact on the system. This is
partly because, for legibility reasons, we scale vectors such that the largest
vector in the plot has a length equal to the distance between neighboring
nominal points.
In panels (\subref{fig:fig1_formose}) and (\subref{fig:product_formose}), the
points and arrows are colored according to the value of the observable (number
of different molecular species created throughout the simulation), while in
panel (\subref{fig:fig2_formose}), points are colored according to the
magnitude of the full, non-projected gradient vector, highlighting regions of
higher and lower sensitivity; no arrows are shown in this panel. Arrows in
(\subref{fig:fig1_formose}) and (\subref{fig:product_formose}) indicate both
the direction and relative magnitude of the local sensitivity, with points
showing very small or zero gradient magnitude appearing without visible arrows.
We can also observe a slight sensitivity with respect to changes along the
keto-enol dimension in the area where this parameter is zero.

\begin{figure}
    \centering
    \begin{subfigure}[b]{0.50\textwidth}
        \centering
        \hspace*{3.5em}%

        \includegraphics[scale=0.5]{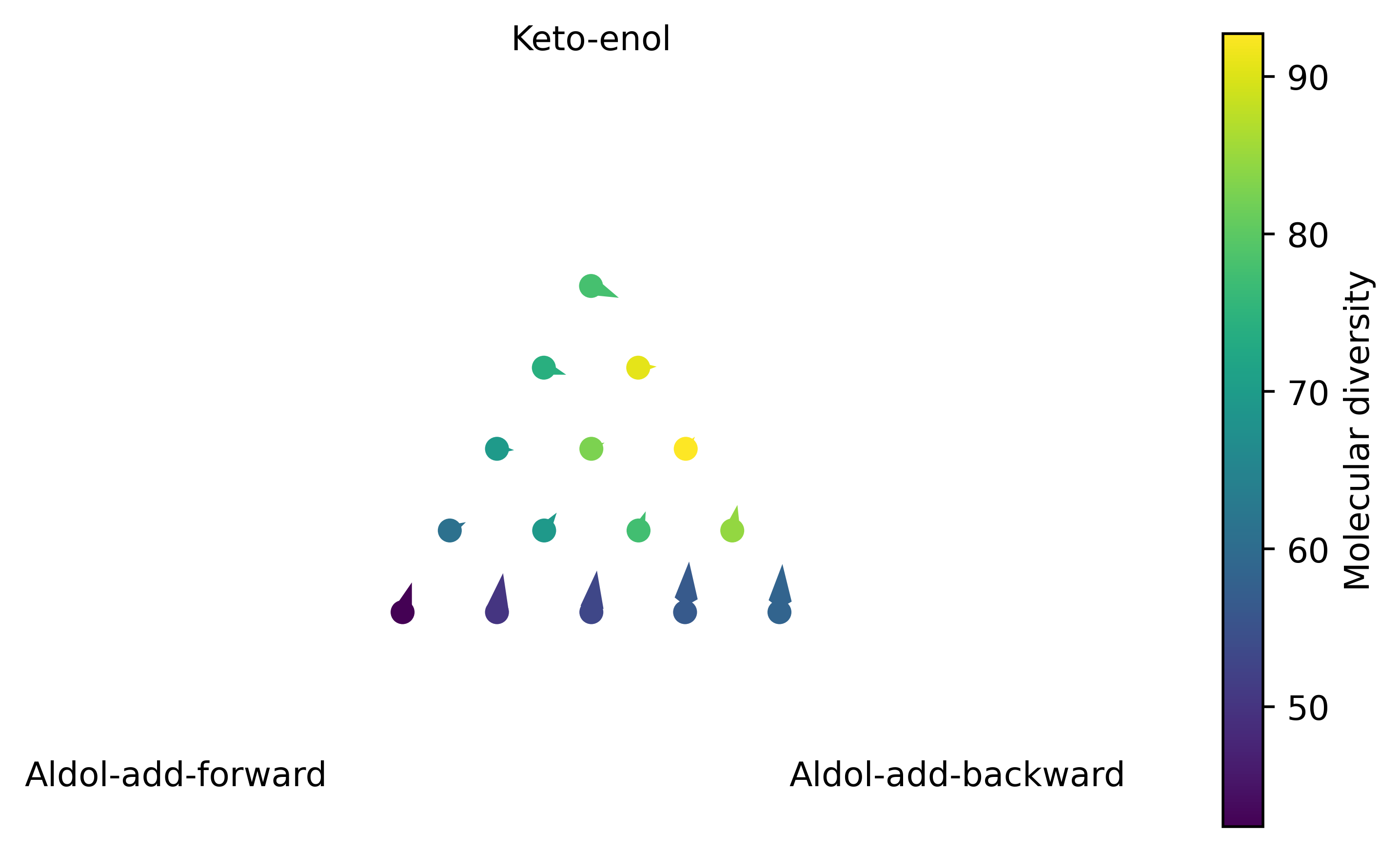}
        \caption{Projected gradient vectors.}
        \label{fig:fig1_formose_nonzero}
    \end{subfigure}%
    \begin{subfigure}[b]{0.50\textwidth}
        \centering
        \hspace*{3.5em}%

        \includegraphics[scale=0.5]{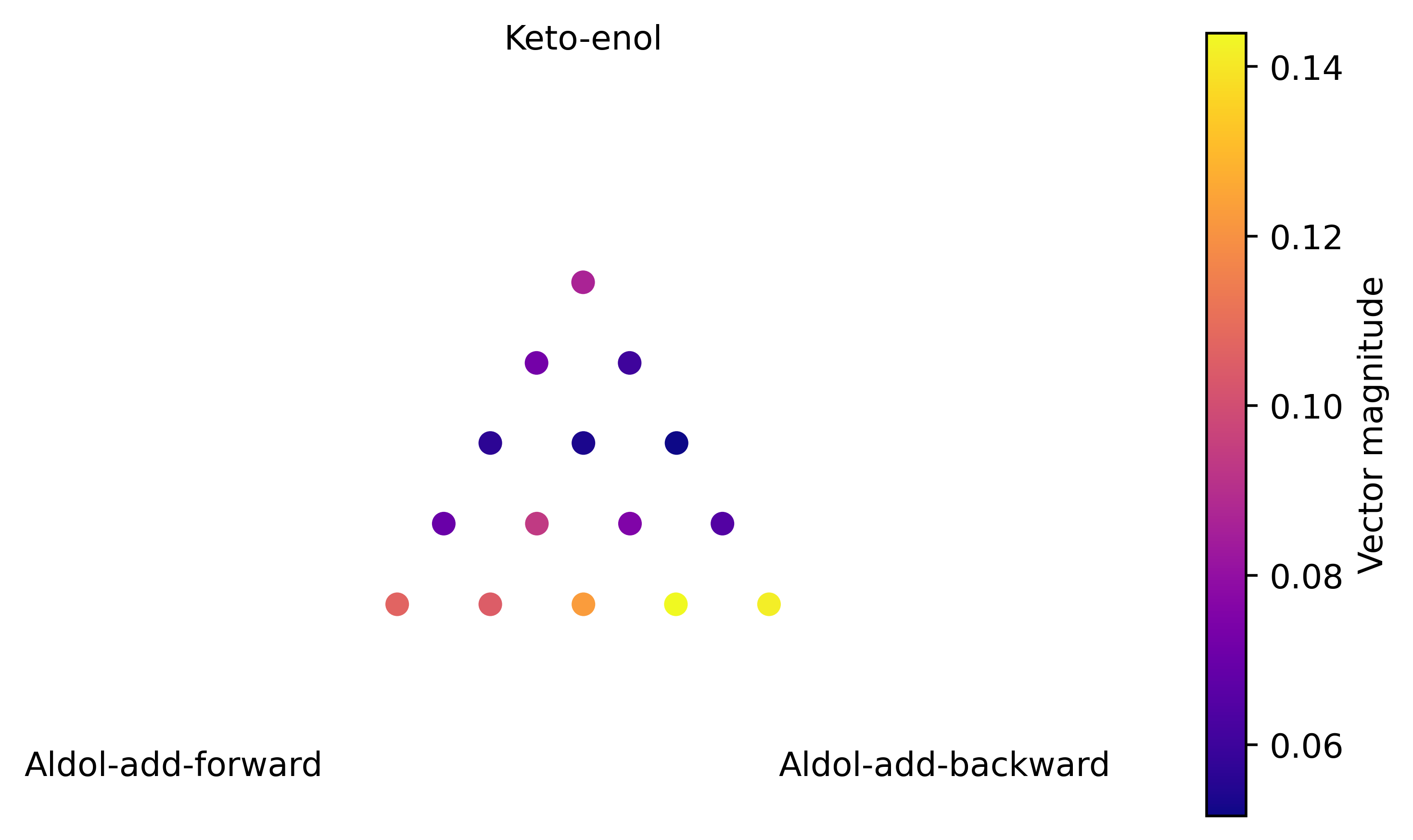}
        \caption{3D magnitude colour map.}
        \label{fig:fig2_formose_nonzero}
    \end{subfigure}

    \begin{subfigure}[b]{\textwidth}

        \includegraphics[width=\linewidth]{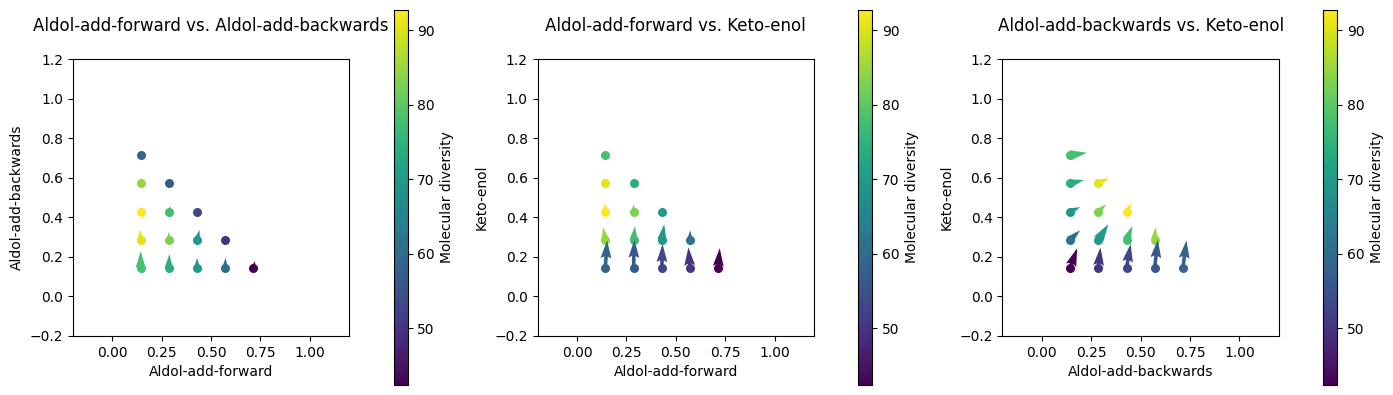}
        \caption{Gradient vectors projected into each of the 3 basic 2D planes
            of the 3D coordinate system.}
        \label{fig:product_formose_nonzero}
    \end{subfigure}
    \caption{Sensitivity analysis of the formose model.
        \subref{fig:fig1_formose_nonzero} Gradient vectors projected onto the
        2-simplex plane.
        Each point represents a sampled parameter set; the arrow at each point
        shows
        the local sensitivity (gradient vector) of the observable with respect
        to all
        three rate constants. The color of each point and vector indicates the
        value of
        the observable (number of different molecular species created
        throughout the simulation). Points with very small
        or zero
        gradient magnitude appear without a visible arrow.
        \subref{fig:fig2_formose_nonzero} Each sampled point colored according
        to the magnitude
        of the full, non-projected gradient vector; no arrows are shown in this
        panel.
        This highlights regions of high and low sensitivity.
        \subref{fig:product_formose_nonzero} Gradient vectors projected onto
        the three basic 2D
        planes of the 3D coordinate system, with colors indicating the
        observable as in
        panel (a). In this new set of plots we have removed points with zero
        coordinates.
        In
        this setup, the observable is more sensitive to changes along the
        keto-enol
        dimension, and to changes in the aldol-addition$\rightarrow$ rate constant
        in second
        place. Highest sensitivity is achieved in the region where the
        keto-enol rate
        constants are small.}
    \label{fig:product-combined_formose_nonzero}
\end{figure}

To investigate further the remaining areas, we remove points with zeroes in
their coordinates (and rescale the remaining vectors), by which we obtain
Figure~\ref{fig:product-combined_formose_nonzero}, which follows the same
visual conventions as the previous figure.
Here, the sensitivity of the system (length of vectors) displays a different
pattern: the system seems sensitive to changes in rule rate constants for
both the aldol-addition$\leftarrow$ and keto-enol rules, but especially for the
keto-enol rules, in the region of space where this parameter takes values
closer to zero.
An interesting maximal point (yellow) can be observed, where the chemical space
is as diverse as it can get in the chosen amount of simulation time.
Lower diversity is in general achieved for low rate constants along the
keto-enol axis, as indicated by the darker purple points.

\paragraph{Global Sensitivity Coefficients.}
We repeat the experiment five times to generate statistics on the global sensitivity
coefficients.
Table~\ref{tab:sens_coeff} provides a summary.
As the concluding remark, the aldol-addition$\rightarrow$ rate constant impacts
the system's diversity the most, followed by both keto-enol rule rate
constants.
\begin{table}
    \centering
    \sisetup{table-format=3.3}
    \begin{tabular}{@{}lSSSSSS@{}}
        \toprule
        Dimension          & {Avg}  & {Median} & {Min}   & {Max}   & {Std Dev}
        \\
        \midrule
        Aldol-add-forward  & 220.722 & 220.8    & 219.664 & 221.538 & 0.714
        \\
        Aldol-add-backward & 59.136  & 58.76    & 57.12   & 61.196  & 1.526
        \\
        Keto-enol          & 127.676 & 127.615  & 125.139 & 129.763 & 1.728
        \\
        \bottomrule
    \end{tabular}
    \caption{Statistics for the sensitivity coefficient for each
        parameter.}
    \label{tab:sens_coeff}
\end{table}
\\

\noindent\textit{Conclusion of the sensitivity analysis:} The diversity of the generated chemical space is most sensitive to changes in the aldol addition forward rate constant, especially in the vicinity of $0$. The second most important parameter is the rate constant for the keto-enol reactions. When points with zero values are removed from the parameter space, interestingly the aldol addition forward rate constant makes the least impact in the observable, being the other two dimensions the most relevant ones.

\subsubsection{Uncertainty Analysis}
\begin{figure}[h!]
    \centering
    \begin{subfigure}[t]{0.32\textwidth}
        \centering
        \includegraphics[width=\textwidth]{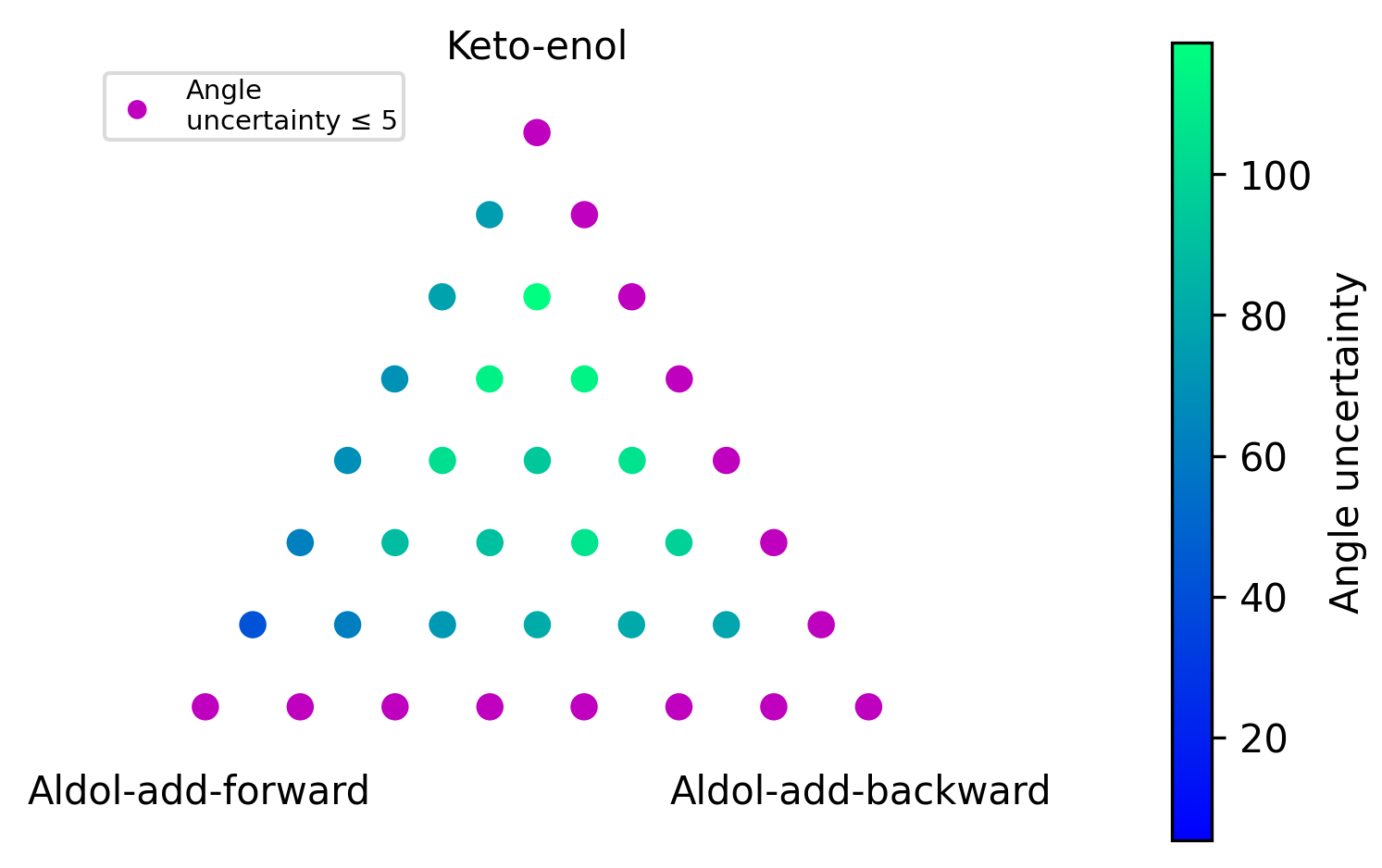}
        \caption{Aldol-add-forward.}
        \label{fig:al-for}
    \end{subfigure}
    \hfill
    \begin{subfigure}[t]{0.32\textwidth}
        \centering
        \includegraphics[width=\textwidth]{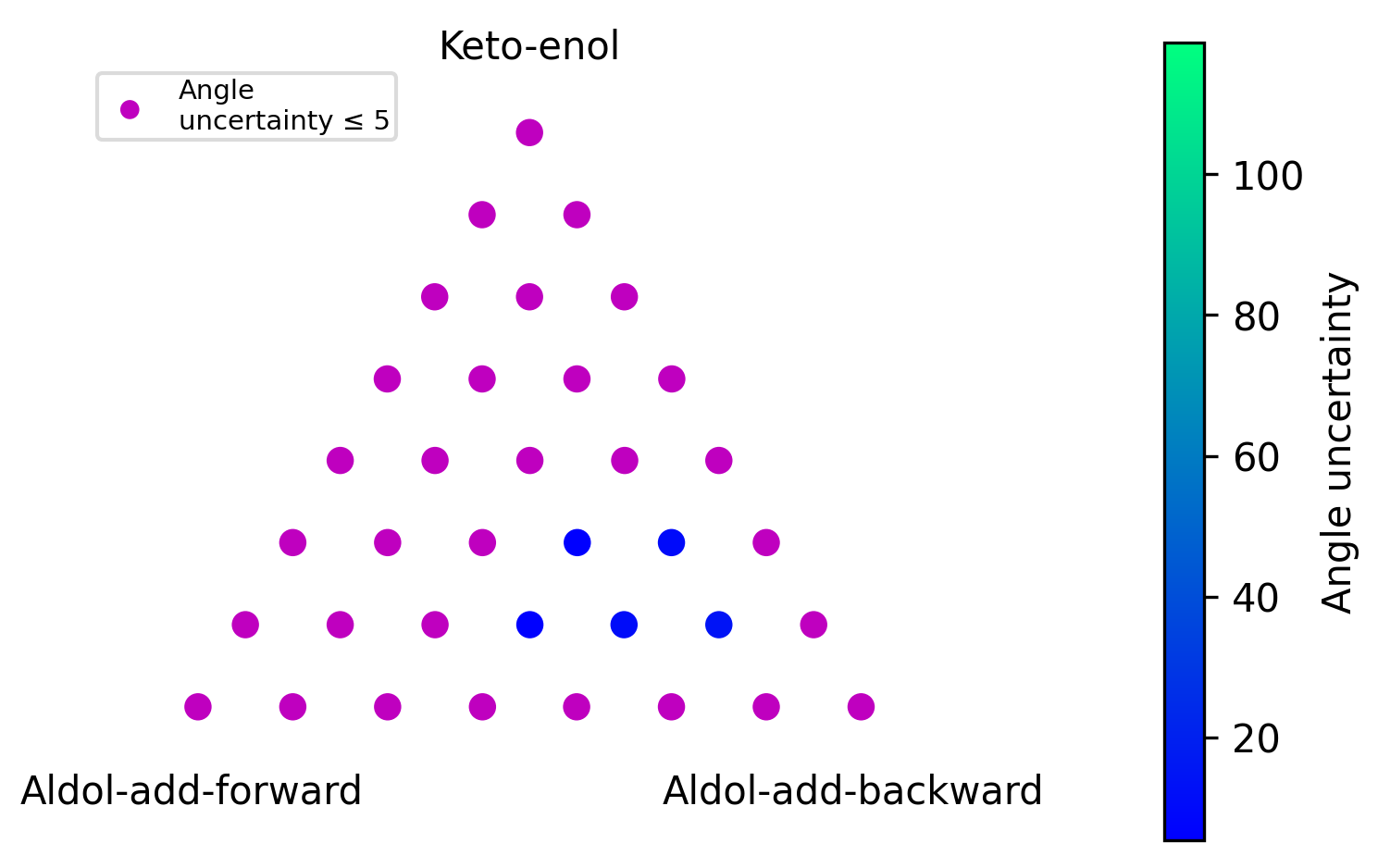}
        \caption{Aldol-add-backward.}
        \label{fig:al-back}
    \end{subfigure}
    \hfill
    \begin{subfigure}[t]{0.32\textwidth}
        \centering
        \includegraphics[width=\textwidth]{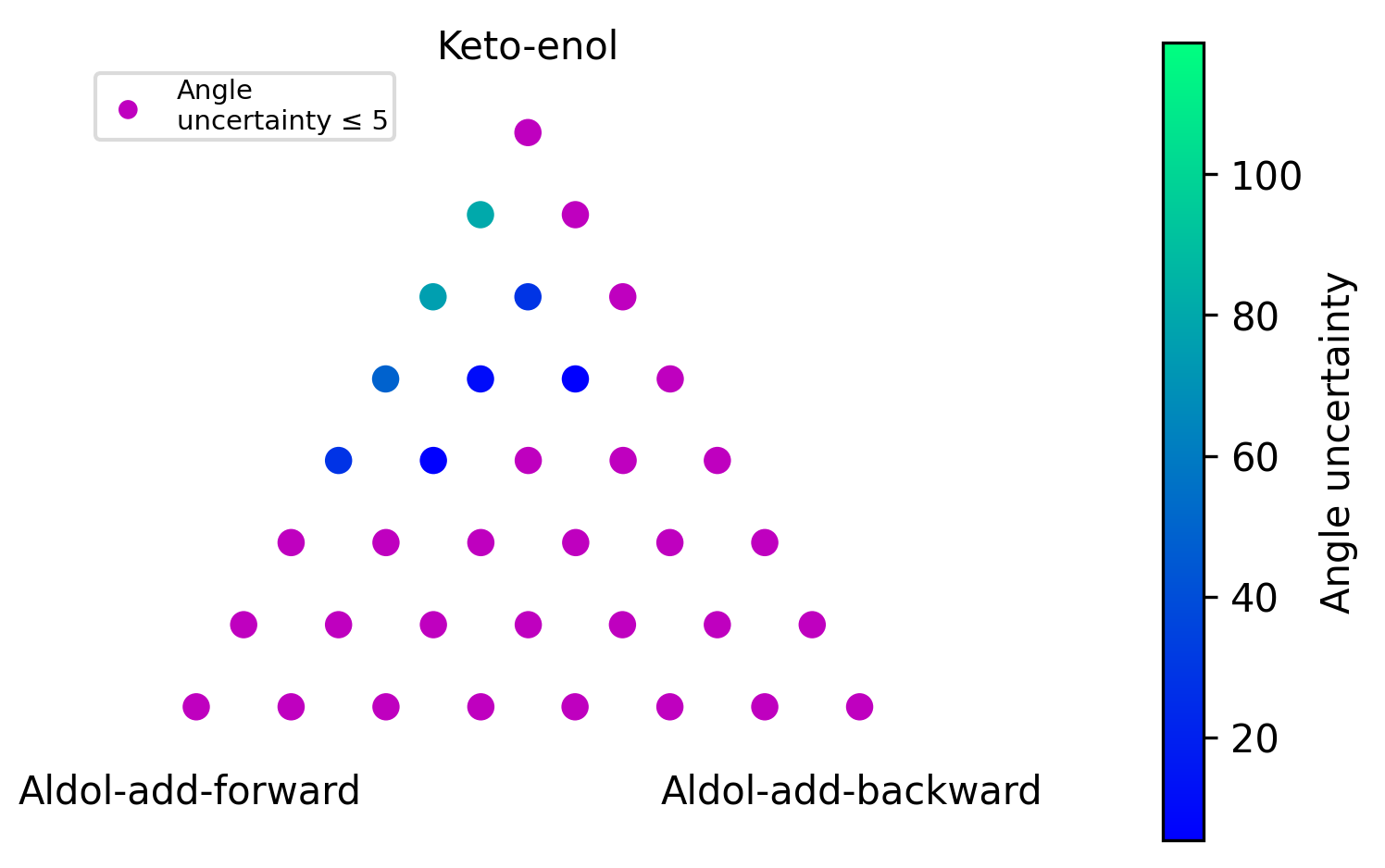}
        \caption{Keto-enol.}
        \label{fig:ketoenol}
    \end{subfigure}

    \caption{Gradient uncertainty across the parameter space. %with a subplot showing the corresponding angle uncertainty at each point, per dimension being perturbed.
      Each subplot considers one perturbed input dimension (i.e., one component of the gradient vector) and shows at each point in parameter space the angle uncertainty of this dimension.
      The average gradient-uncertainty angle across all points is for each dimension as follows:
    \subref{fig:al-for} aldol addition forward (50.167), 
    \subref{fig:al-back} aldol addition backward (2.736), 
    \subref{fig:ketoenol} keto-enol (8.910).
    }
    \label{fig:formose_uncertainty}
\end{figure}

Table~\ref{tab:summary_formose} presents a summary of the angles of gradient uncertainty obtained in this experiment.
We observe an average angle uncertainty of approximately \qty{20.6}{\degree}, with a substantial standard deviation that indicates non-uniform accuracy of the gradient estimates across the parameter space and input dimensions.
This spread reflects the variability in estimated gradient directions when perturbing along different dimensions: some regions and dimensions yield more consistent gradient directions (low angle uncertainty), while others are less well-constrained (high angle uncertainty), see Figure \ref{fig:formose_uncertainty} for details.
In particular, high angle uncertainty highlights perturbation directions where gradient estimates are less reliable, which could suggest the need for additional simulations in these areas to improve estimate accuracy.
\begin{table}
    \centering
    \sisetup{table-format=2.3}
    \begin{tabular}{@{}lSS[table-format=1.3]S[table-format=3.3]S@{}}
        \toprule
                          & {Avg} & {Min} & {Max}   & {Std Dev} \\
        \midrule
        Gradient-Uncertainty Angle & 20.604 & 0.000 & 118.630 & 35.463    \\
        \bottomrule
    \end{tabular}
    \caption{Summary statistics for the gradient-uncertainty angle (across all nominal points and all three input dimensions), measured
        in
        degrees. High variation across the input space can be seen.}
    \label{tab:summary_formose}
\end{table}
 
The aldol-addition$\rightarrow$ rule not only contributes the most to system
sensitivity but also exhibits the highest average angle uncertainty. This
indicates that when perturbing along this dimension, the range of possible gradient
directions is relatively wide, making gradient estimates less consistent. In
contrast, the remaining dimensions yield lower angle uncertainty, indicating
more reliable gradient estimates along those dimensions. 
\\
% Angle uncertainty thus quantifies the variability of the gradient estimates
% along each perturbation direction, helping to identify which axes produce
% consistent gradient directions and which are more susceptible to estimation
% noise.

%
% 
\noindent\textit{Conclusion of the uncertainty analysis:} The aldol addition backward rate constant yields the most reliable gradient estimates, followed closely by the keto-enol parameter. 

\paragraph{Extended Uncertainty Analysis: Different Runtime Limits.}

\begin{figure}
    \centering

    \includegraphics[width=0.6\linewidth]{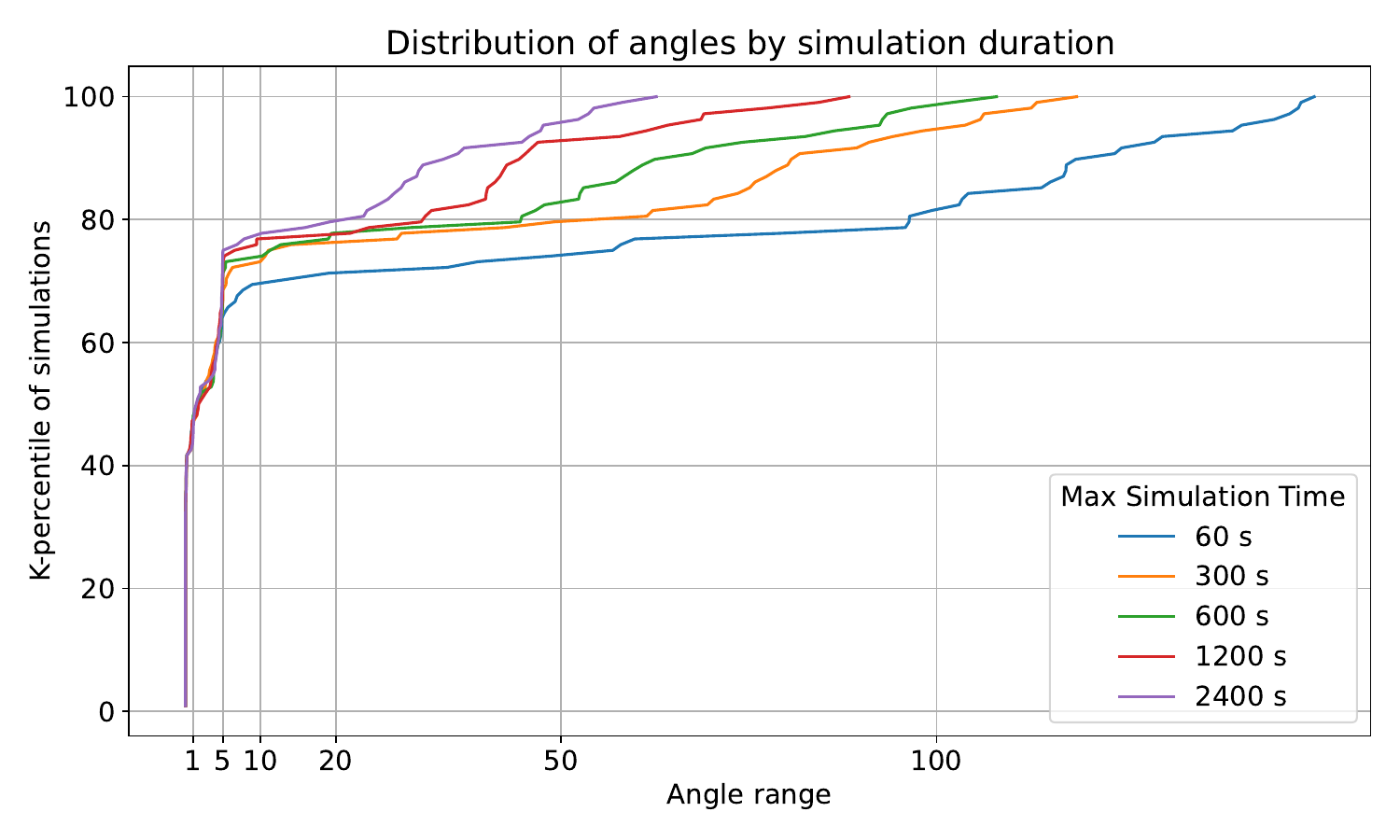}
    \caption{Percentile distribution of the number of times each indicated
        angle range was reached, across all nominal points and perturbations.
        Each
        colored line corresponds to a different runtime limit used when
        simulating each
        nominal–perturbation pair.}
    \label{fig:f-angle}
\end{figure}

In Figure~\ref{fig:f-angle}, we extend the statistics of
Table~\ref{tab:summary_formose} on angles by plotting the number of times (in
percent across all nominal points and perturbation experiments) that we
achieved a given angle bound when we conducted the experiments with different
runtime limits.
It can be observed that, in approximately \qty{68}{\percent} of the cases, we
reached the \qty{5}{\degree} angle bound when the time limit is equal to or
greater than $300$ seconds wall clock time. %(used in the experiments above).
As the available simulation time increases, we observe a moderate increase in
the percentage of cases where we obtained a bound of \qty{5}{\degree} or less,
ranging from approximately \qty{64}{\percent} (blue) to \qty{75}{\percent}
(purple). Similarly to the example above, the maximum angle obtained for each
time limit varies widely, ranging approximately from \qty{63}{\degree} (purple)
to \qty{150}{\degree} (blue).

\subsection{Further Examples of Observables}

Note that a strong aspect of our methodology is the flexibility to work with
many different observables. The possibilities are wide: as previously
mentioned, in MØD we can retrieve the simulation trace of events, where each
rule application (and chemical reaction instantiated by this rule) is
specified. \textit{Dynamic} information, such as event times, molecular species
at each simulation step, and species counts over time, is accessible. From
this, one can define dynamic observables such as the time until a certain
molecule appears or until the first application of a specific rule (more
generally, the time to an event of interest), the distribution of molecular
counts over time, or the total distribution of rule applications. Additionally,
\textit{structural} information is also available, including molecular
structures down to the atomic level, which allows for defining observables such
as identifying structural motifs, finding specific molecules, or simply
counting atoms like carbons. Chemical network properties, such as the size of
the network, like in our formose example, are also available. Taken together,
this combination of dynamic and structural information supports the
construction of a broad and flexible set of observables.

As concrete examples, for the formose system in subsection
\nameref{Subsec:Formose} we experimented with the following additional
observables:

\begin{itemize}
    \item Count of pentoses: we created a function that analyzes the chemical
          structure of each species, checks whether it is a carbohydrate, and
          counts the
          number of carbons. We can analyze variations in pentose count
          throughout the
          simulation, or focus on the amount obtained at the final simulation
          time.
    \item Count of a specific molecular pattern: for example, structures with
          branched carbons or formose-related chemical motifs. We used MØD's
          monomorphism
          check to search for specific substructures in our chemical space.
          Concretely,
          the following SMILES patterns: \verb|[C][C]([C])([C])|, and
          \verb|OCC(O)C(O)C=O|.
    \item Count of specific molecules: we analyzed the sensitivity to changes
          in the counts of both glycolaldehyde and formaldehyde at the end of
          the
          simulation.
\end{itemize}
The sensitivity analysis conclusions were essentially the same as those already
reported in subsection \nameref{Subsec:Formose}, hence, for reasons of space,
we omit plots for those additional experiments.

\subsection{Technical Details}
Computations were performed on a laptop with a 13th Gen Intel Core i7-13700H
processor with 32 GB of RAM,
using Python 3.10 and MØD version 0.16.1.77. Note that the methodology allows
for a high degree of parallelism; here, we parallelized the simulations
conducted for the nominal point and the perturbation point. The total runtime
for conducting one experiment of the Michaelis-Menten system, allowing a
runtime limit of $300$ seconds per simulated nominal–perturbation, and
evaluating 144 input points (nominal and perturbed), took 3 hours and 2
minutes.
For the formose system, for the same runtime limit and the same number of
points, 3 hours and 16 minutes.

%%%%%%%%%%%%%%%%%%%%%%%%%%%%%%%%%%%%%%%%%%%%%%%%%%%%%%%%%%%%%%%%%%%%%%%%%%%%%%%
\section{Conclusion}\label{Sec:Conclusion}
%%%%%%%%%%%%%%%%%%%%%%%%%%%%%%%%%%%%%%%%%%%%%%%%%%%%%%%%%%%%%%%%%%%%%%%%%%%%%%%
In the context of stochastic chemical systems, many global techniques
demand extensive simulation runs to reduce noise-related
uncertainty. This can become computationally prohibitive if each
simulation is expensive or if the parameter space must be densely
sampled to capture interactions.
Because rule-based models often have a relatively small set of core
parameters (even if they generate a combinatorially large network of
species on-the-fly), a local gradient-based approach across multiple
parameter points can be both feasible and illuminating. By adaptively
increasing simulation repetitions until reaching a desired level of gradient
uncertainty, the precision of the output is systematically controlled, and the use of computation time is optimized.
In addition, limiting the analysis to a small parameter set simplifies result
visualization and further helps keeping computational demands down. The proposed
method retains the simplicity of local, derivative-based analysis while
offering a moderately global perspective by sampling parameter space more
broadly—especially suitable for rule-based and combinatorially complex chemical
systems. This synergy allows for straightforward “black-box” repeated
simulations, customized observables, and effective exploration of the
sensitivity across the parameter space, yielding practical insights into model
robustness and parameter importance.

Interesting directions for future work include incorporating established variance-reduction techniques, such as CRN and CRP, to further reduce the variance of gradient estimators. Additionally, when prior information on the simulated observables is available, the use of credible intervals for gradient estimation could be employed, allowing physical knowledge to be incorporated into the uncertainty of the gradients\cite{savara2020chekipeuq}.
%Finally, methods that accelerate individual simulation runs could be integrated to further improve overall computational efficiency.
Finally, we note that a key idea of our methodology, namely reducing the number of simulations in way adaptive to the stochasticity of each point in parameter space, is completely orthogonal to any efforts on reducing the time for a single simulation and thus has good potential for synergy with any existing general or domain-specific methods for accelerating stochastic simulations.

%%%%%%%%%%%%%%%%%%%%%%%%%%%%%%%%%%%%%%%%%%%%%%%%%%%%%%%%%%%%%%%%%%%%%%%%%%%%%%%
\section{Data and Software Availability}
%%%%%%%%%%%%%%%%%%%%%%%%%%%%%%%%%%%%%%%%%%%%%%%%%%%%%%%%%%%%%%%%%%%%%%%%%%%%%%%
All data and code supporting the findings of this study are openly available in
the GitHub repository:
\url{https://github.com/erihm/sensitivity_analysis_stochastic_chemical_systems}.

\section{Acknowledgement}

    This project has received funding from the European Union's Horizon Europe
    Doctoral Network programme under the Marie Skłodowska-Curie grant agreement
    No. 101072930 (TACsy – Training Alliance for Computational Systems Chemistry).


\bibliographystyle{plain}
\bibliography{bibliography}

@misc{machado2025rulebasedgillespiesimulationchemical,
      title={Rule-Based Gillespie Simulation of Chemical Systems}, 
      author={Erika M. Herrera Machado and Jakob L. Andersen and Rolf Fagerberg and Christoph Flamm and Daniel Merkle and Peter F. Stadler},
      year={2025},
      eprint={2509.01504},
      archivePrefix={arXiv},
      primaryClass={q-bio.MN},
      url={https://arxiv.org/abs/2509.01504}, 
}

@article{andersen2014generic,
  title={Generic strategies for chemical space exploration},
  author={Andersen, Jakob L and Flamm, Christoph and Merkle, Daniel and Stadler, Peter F},
  journal={International journal of computational biology and drug design},
  volume={7},
  number={2-3},
  pages={225--258},
  year={2014},
  publisher={Inderscience Publishers Ltd}
}

@article{rabitz1983sensitivity,
  title={Sensitivity analysis in chemical kinetics},
  author={Rabitz, Herschel and Kramer, Mark and Dacol, D},
  journal={Annual review of physical chemistry},
  volume={34},
  number={1},
  pages={419--461},
  year={1983},
  publisher={Annual Reviews 4139 El Camino Way, PO Box 10139, Palo Alto, CA 94303-0139, USA}
}

@article{kwak2017central,
  title={Central limit theorem: the cornerstone of modern statistics},
  author={Kwak, Sang Gyu and Kim, Jong Hae},
  journal={Korean journal of anesthesiology},
  volume={70},
  number={2},
  pages={144--156},
  year={2017},
  publisher={The Korean Society of Anesthesiologists}
}

@book{ross2017introductory,
  title={Introductory statistics},
  author={Ross, Sheldon M},
  year={2017},
  publisher={Academic Press}
}

@book{bertsekas2008introduction,
  title={Introduction to probability},
  author={Bertsekas, Dimitri and Tsitsiklis, John N},
  volume={1},
  year={2008},
  publisher={Athena Scientific}
}

@inproceedings{andersen2016software,
  title={A software package for chemically inspired graph transformation},
  author={Andersen, Jakob L and Flamm, Christoph and Merkle, Daniel and Stadler, Peter F},
  booktitle={Graph Transformation: 9th International Conference, ICGT 2016, in Memory of Hartmut Ehrig, Held as Part of STAF 2016, Vienna, Austria, July 5-6, 2016, Proceedings 9},
  pages={73--88},
  year={2016},
  organization={Springer}
}

@book{burden2001,
author={Burden,Richard L. and Faires,J. D.},
year={2001},
title={Numerical analysis},
publisher={Brooks/Cole},
address={Pacific Grove, CA},
edition={7th},
isbn={9780534382162;9780534382179;0534382177;0534382169;},
language={English},
}

@book{fu2015stochastic,
  title={Stochastic gradient estimation},
  author={Fu, Michael C},
  year={2015},
  publisher={Springer}
}

@book{schuster2016stochasticity,
  title={Stochasticity in processes},
  author={Schuster, Peter},
  year={2016},
  publisher={Springer}
}

@article{michaelis1913kinetics,
  title={The kinetics of the inversion effect},
  author={Michaelis, Leonor and Menten, Maude L},
  journal={Biochem. Z},
  volume={49},
  pages={333--369},
  year={1913}
}

@article{damiani2013parameter,
  title={Parameter sensitivity analysis of stochastic models: Application to catalytic reaction networks},
  author={Damiani, Chiara and Filisetti, Alessandro and Graudenzi, Alex and Lecca, Paola},
  journal={Computational biology and chemistry},
  volume={42},
  pages={5--17},
  year={2013},
  publisher={Elsevier}
}

@article{saltelli2005sensitivity,
  title={Sensitivity analysis for chemical models},
  author={Saltelli, Andrea and Ratto, Marco and Tarantola, Stefano and Campolongo, Francesca},
  journal={Chemical reviews},
  volume={105},
  number={7},
  pages={2811--2828},
  year={2005},
  publisher={ACS Publications}
}

@article{plyasunov2007efficient,
  title={Efficient stochastic sensitivity analysis of discrete event systems},
  author={Plyasunov, Sergey and Arkin, Adam P},
  journal={Journal of Computational Physics},
  volume={221},
  number={2},
  pages={724--738},
  year={2007},
  publisher={Elsevier}
}

@book{nocedal1999numerical,
  title={Numerical optimization},
  author={Nocedal, Jorge and Wright, Stephen J},
  year={1999},
  publisher={Springer}
}

@article{bartholomay1958stochastic,
  title={Stochastic models for chemical reactions: I. Theory of the unimolecular reaction process},
  author={Bartholomay, Anthony F},
  journal={The bulletin of mathematical biophysics},
  volume={20},
  pages={175--190},
  year={1958},
  publisher={Springer}
}

@article{gillespie1976general,
  title={A general method for numerically simulating the stochastic time evolution of coupled chemical reactions},
  author={Gillespie, Daniel T},
  journal={Journal of computational physics},
  volume={22},
  number={4},
  pages={403--434},
  year={1976},
  publisher={Elsevier}
}

@inproceedings{danos2007rule,
  title={Rule-based modelling of cellular signalling},
  author={Danos, Vincent and Feret, J{\'e}r{\^o}me and Fontana, Walter and Harmer, Russell and Krivine, Jean},
  booktitle={International conference on concurrency theory},
  pages={17--41},
  year={2007},
  organization={Springer}
}

@article{benko2003graph,
  title={A graph-based toy model of chemistry},
  author={Benk{\"o}, Gil and Flamm, Christoph and Stadler, Peter F},
  journal={Journal of Chemical Information and Computer Sciences},
  volume={43},
  number={4},
  pages={1085--1093},
  year={2003},
  publisher={ACS Publications}
}

@article{rathinam2010efficient,
  title={Efficient computation of parameter sensitivities of discrete stochastic chemical reaction networks},
  author={Rathinam, Muruhan and Sheppard, Patrick W and Khammash, Mustafa},
  journal={The Journal of chemical physics},
  volume={132},
  number={3},
  year={2010},
  publisher={AIP Publishing}
}

@article{komorowski2011sensitivity,
  title={Sensitivity, robustness, and identifiability in stochastic chemical kinetics models},
  author={Komorowski, Micha{\l} and Costa, Maria J and Rand, David A and Stumpf, Michael PH},
  journal={Proceedings of the National Academy of Sciences},
  volume={108},
  number={21},
  pages={8645--8650},
  year={2011},
  publisher={National Acad Sciences}
}

@article{gunawan2005sensitivity,
  title={Sensitivity analysis of discrete stochastic systems},
  author={Gunawan, Rudiyanto and Cao, Yang and Petzold, Linda and Doyle, Francis J},
  journal={Biophysical journal},
  volume={88},
  number={4},
  pages={2530--2540},
  year={2005},
  publisher={Elsevier}
}

@article{chylek2014rule,
  title={Rule-based modeling: a computational approach for studying biomolecular site dynamics in cell signaling systems},
  author={Chylek, Lily A and Harris, Leonard A and Tung, Chang-Shung and Faeder, James R and Lopez, Carlos F and Hlavacek, William S},
  journal={Wiley Interdisciplinary Reviews: Systems Biology and Medicine},
  volume={6},
  number={1},
  pages={13--36},
  year={2014},
  publisher={Wiley Online Library}
}

@article{sobol1990sensitivity,
  title={On sensitivity estimation for nonlinear mathematical models},
  author={Sobol', Il'ya Meerovich},
  journal={Matematicheskoe modelirovanie},
  volume={2},
  number={1},
  pages={112--118},
  year={1990},
  publisher={Russian Academy of Sciences, Branch of Mathematical Sciences}
}

@article{costanza1981stochastic,
  title={Stochastic sensitivity analysis in chemical kinetics},
  author={Costanza, Vicente and Seinfeld, John H},
  journal={The Journal of chemical physics},
  volume={74},
  number={7},
  pages={3852--3858},
  year={1981},
  publisher={AIP Publishing}
}

@article{morshed2017efficient,
  title={An efficient finite-difference strategy for sensitivity analysis of stochastic models of biochemical systems},
  author={Morshed, Monjur and Ingalls, Brian and Ilie, Silvana},
  journal={Biosystems},
  volume={151},
  pages={43--52},
  year={2017},
  publisher={Elsevier}
}

@article{anderson2012efficient,
  title={An efficient finite difference method for parameter sensitivities of continuous time Markov chains},
  author={Anderson, David F},
  journal={SIAM Journal on Numerical Analysis},
  volume={50},
  number={5},
  pages={2237--2258},
  year={2012},
  publisher={SIAM}
}

@article{turanyi1990sensitivity,
  title={Sensitivity analysis of complex kinetic systems. Tools and applications},
  author={Tur{\'a}nyi, Tam{\'a}s},
  journal={Journal of mathematical chemistry},
  volume={5},
  number={3},
  pages={203--248},
  year={1990},
  publisher={Springer}
}

@article{morio2011global,
  title={Global and local sensitivity analysis methods for a physical system},
  author={Morio, J{\'e}r{\^o}me},
  journal={European journal of physics},
  volume={32},
  number={6},
  pages={1577},
  year={2011},
  publisher={IOP Publishing}
}

@article{cukier1978nonlinear,
  title={Nonlinear sensitivity analysis of multiparameter model systems},
  author={Cukier, RI and Levine, HB and Shuler, KE},
  journal={Journal of computational physics},
  volume={26},
  number={1},
  pages={1--42},
  year={1978},
  publisher={Elsevier}
}

@article{cukier1975study,
  title={Study of the sensitivity of coupled reaction systems to uncertainties in rate coefficients. III. Analysis of the approximations},
  author={Cukier, RI and Schaibly, JH and Shuler, Kurt E},
  journal={The Journal of Chemical Physics},
  volume={63},
  number={3},
  pages={1140--1149},
  year={1975},
  publisher={AIP Publishing}
}

@article{morris1991factorial,
  title={Factorial sampling plans for preliminary computational experiments},
  author={Morris, Max D},
  journal={Technometrics},
  volume={33},
  number={2},
  pages={161--174},
  year={1991},
  publisher={Taylor \& Francis}
}

@article{atherton1975statistical,
  title={On the statistical sensitivity analysis of models for chemical kinetics},
  author={Atherton, RW and Schainker, RB and Ducot, ER},
  journal={AIChE Journal},
  volume={21},
  number={3},
  pages={441--448},
  year={1975},
  publisher={Wiley Online Library}
}

@article{miller1983sensitivity,
  title={Sensitivity analysis and parameter estimation in dynamic modeling of chemical kinetics},
  author={Miller, David and Frenklach, Michael},
  journal={International Journal of Chemical Kinetics},
  volume={15},
  number={7},
  pages={677--696},
  year={1983},
  publisher={Wiley Online Library}
}

@article{zador2006local,
  title={Local and global uncertainty analysis of complex chemical kinetic systems},
  author={Z{\'a}dor, Judit and Zsely, I Gy and Tur{\'a}nyi, Tam{\'a}s},
  journal={Reliability Engineering \& System Safety},
  volume={91},
  number={10-11},
  pages={1232--1240},
  year={2006},
  publisher={Elsevier}
}

@incollection{degasperi2008sensitivity,
  title={Sensitivity analysis of stochastic models of bistable biochemical reactions},
  author={Degasperi, Andrea and Gilmore, Stephen},
  booktitle={International School on Formal Methods for the Design of Computer, Communication and Software Systems},
  pages={1--20},
  year={2008},
  publisher={Springer}
}

@article{sheppard2013spsens,
  title={SPSens: a software package for stochastic parameter sensitivity analysis of biochemical reaction networks},
  author={Sheppard, Patrick W and Rathinam, Muruhan and Khammash, Mustafa},
  journal={Bioinformatics},
  volume={29},
  number={1},
  pages={140--142},
  year={2013},
  publisher={Oxford University Press}
}

@article{savara2020chekipeuq,
  title={CheKiPEUQ intro 1: Bayesian parameter estimation considering uncertainty or error from both experiments and theory},
  author={Savara, Aditya and Walker, Eric A},
  journal={ChemCatChem},
  volume={12},
  number={21},
  pages={5385--5400},
  year={2020},
  publisher={Wiley Online Library}
}

\end{document}